\documentclass[twocolumn]{aastex7}
\usepackage{url}
\usepackage{amsmath}
\usepackage{multirow}
\usepackage{booktabs}
\usepackage{hyperref}
\usepackage{latexsym}
\received{}
\revised{}
\accepted{}
\published{}
\shorttitle{SN~2024aecx}
\shortauthors{X. Zou \& B. Kumar et al.}
\begin{document}

\title{SN~2024aecx: A double-peaked rapidly evolving Type IIb supernova at 11 Mpc}
\correspondingauthor{Brajesh Kumar} 
\email{brajesh@ynu.edu.cn, brajesharies@gmail.com}

\author[0009-0006-5847-9271]{Xingzhu Zou}
\email{xingzhuzou@mail.ynu.edu.cn}
\affiliation{South-Western Institute for Astronomy Research, Yunnan University, Kunming, Yunnan 650504, People's Republic of China}
\affiliation{Yunnan Key Laboratory of Survey Science, Yunnan University, Kunming, Yunnan 650500, People's Republic of China}

\author[0000-0001-7225-2475]{Brajesh Kumar}
\email{brajesh@ynu.edu.cn, brajesharies@gmail.com}
\affiliation{South-Western Institute for Astronomy Research, Yunnan University, Kunming, Yunnan 650504, People's Republic of China}
\affiliation{Yunnan Key Laboratory of Survey Science, Yunnan University, Kunming, Yunnan 650500, People's Republic of China}

\author[0000-0002-0525-0872]{Rishabh Singh Teja}
\email{rsteja001@gmail.com}
\affiliation{Indian Institute of Astrophysics, II Block, Koramangala, Bengaluru-560034, Karnataka, India}
\affiliation{Tsung-Dao Lee Institute, Shanghai Jiao Tong University, No.1 Lisuo Road, Pudong New Area, Shanghai, People's Republic of China}

\author[0000-0002-6688-0800]{D. K. Sahu}
\email{dks@iiap.res.in}
\affiliation{Indian Institute of Astrophysics, II Block, Koramangala, Bengaluru-560034, Karnataka, India}

\author[0009-0000-4068-1320]{Xinlei Chen}
\email{xlchen@stu.ynu.edu.cn}
\affiliation{South-Western Institute for Astronomy Research, Yunnan University, Kunming, Yunnan 650504, People's Republic of China}
\affiliation{Yunnan Key Laboratory of Survey Science, Yunnan University, Kunming, Yunnan 650500, People's Republic of China}

\author[0000-0003-2091-622X]{Avinash Singh}
\email{avinash21292@gmail.com}
\affiliation{The Oskar Klein Centre, Department of Astronomy, Stockholm University, AlbaNova, SE-10691 Stockholm, Sweden}
\affiliation{Hiroshima Astrophysical Science Center, Hiroshima University, Higashi-Hiroshima, Hiroshima 739-8526, Japan}

\author[0000-0003-2240-7031]{Weikang Lin}
\email{weikanglin@ynu.edu.cn}
\affiliation{South-Western Institute for Astronomy Research, Yunnan University, Kunming, Yunnan 650504, People's Republic of China}
\affiliation{Yunnan Key Laboratory of Survey Science, Yunnan University, Kunming, Yunnan 650500, People's Republic of China}

\author[0000-0003-0394-1298]{Xiangkun Liu}
\email{liuxk@ynu.edu.cn}
\affiliation{South-Western Institute for Astronomy Research, Yunnan University, Kunming, Yunnan 650504, People's Republic of China}
\affiliation{Yunnan Key Laboratory of Survey Science, Yunnan University, Kunming, Yunnan 650500, People's Republic of China}

\author[0000-0002-0409-5719]{Dezi Liu}
\email{adzliu@ynu.edu.cn}
\affiliation{South-Western Institute for Astronomy Research, Yunnan University, Kunming, Yunnan 650504, People's Republic of China}
\affiliation{Yunnan Key Laboratory of Survey Science, Yunnan University, Kunming, Yunnan 650500, People's Republic of China}

\author[0009-0007-9727-7792]{Hrishav Das}
\email{hrishav.das@iiap.res.in}
\affiliation{Indian Institute of Astrophysics, II Block, Koramangala, Bengaluru-560034, Karnataka, India}

\author[0000-0001-6706-2749]{Mridweeka Singh}
\email{yashasvi04@gmail.com}
\affiliation{Indian Institute of Astrophysics, II Block, Koramangala, Bengaluru-560034, Karnataka, India}

\author[0000-0003-3533-7183]{G. C. Anupama}
\email{gca@iiap.res.in}
\affiliation{Indian Institute of Astrophysics, II Block, Koramangala, Bengaluru-560034, Karnataka, India}

\author[0009-0002-7625-2653]{Yu Pan}
\email{yupan0304@163.com}
\affiliation{South-Western Institute for Astronomy Research, Yunnan University, Kunming, Yunnan 650504, People's Republic of China}
\affiliation{Yunnan Key Laboratory of Survey Science, Yunnan University, Kunming, Yunnan 650500, People's Republic of China}

\author[0000-0002-8109-7152]{Guowang Du}
\email{dugking@ynu.edu.cn}
\affiliation{South-Western Institute for Astronomy Research, Yunnan University, Kunming, Yunnan 650504, People's Republic of China}
\affiliation{Yunnan Key Laboratory of Survey Science, Yunnan University, Kunming, Yunnan 650500, People's Republic of China}

\author[0000-0001-5737-6445]{Helong Guo}
\email{hlguo@ynu.edu.cn}
\affiliation{South-Western Institute for Astronomy Research, Yunnan University, Kunming, Yunnan 650504, People's Republic of China}
\affiliation{Yunnan Key Laboratory of Survey Science, Yunnan University, Kunming, Yunnan 650500, People's Republic of China}

\author[0009-0005-8762-0871]{Tao Wang}
\email{wangtao@itc.ynu.edu.cn}
\affiliation{South-Western Institute for Astronomy Research, Yunnan University, Kunming, Yunnan 650504, People's Republic of China}
\affiliation{Yunnan Key Laboratory of Survey Science, Yunnan University, Kunming, Yunnan 650500, People's Republic of China}

\author[0009-0003-6936-7548]{Xufeng Zhu}
\email{xufengzhu@ynu.edu.cn}
\affiliation{South-Western Institute for Astronomy Research, Yunnan University, Kunming, Yunnan 650504, People's Republic of China}
\affiliation{Yunnan Key Laboratory of Survey Science, Yunnan University, Kunming, Yunnan 650500, People's Republic of China}

\author[0000-0002-8296-2590]{Jujia Zhang}
\email{jujia@ynao.ac.cn}
\affiliation{Yunnan Observatories (YNAO), Chinese Academy of Sciences (CAS), Kunming, 650216, People's Republic of China}
\affiliation{International Centre of Supernovae, Yunnan Key Laboratory, Kunming 650216, People's Republic of China}
\affiliation{Key Laboratory for the Structure and Evolution of Celestial Objects, CAS, Kunming, 650216, China}

\author[0009-0006-1010-1325]{Yuan Fang}
\email{fangyuan@ynu.edu.cn}
\affiliation{South-Western Institute for Astronomy Research, Yunnan University, Kunming, Yunnan 650504, People's Republic of China}
\affiliation{Yunnan Key Laboratory of Survey Science, Yunnan University, Kunming, Yunnan 650500, People's Republic of China}

\author[0000-0001-5561-2010]{Chenxu Liu}
\email{cxliu@ynu.edu.cn}
\affiliation{South-Western Institute for Astronomy Research, Yunnan University, Kunming, Yunnan 650504, People's Republic of China}
\affiliation{Yunnan Key Laboratory of Survey Science, Yunnan University, Kunming, Yunnan 650500, People's Republic of China}

\author[0000-0002-6252-3750]{Kaushik Chatterjee}
\email{mails.kc.physics@gmail.com}
\affiliation{South-Western Institute for Astronomy Research, Yunnan University, Kunming, Yunnan 650504, People's Republic of China}
\affiliation{Yunnan Key Laboratory of Survey Science, Yunnan University, Kunming, Yunnan 650500, People's Republic of China}

\author[0000-0001-6374-8313]{Yuan-Pei Yang}
\email{ypyang@ynu.edu.cn}
\affiliation{South-Western Institute for Astronomy Research, Yunnan University, Kunming, Yunnan 650504, People's Republic of China}
\affiliation{Yunnan Key Laboratory of Survey Science, Yunnan University, Kunming, Yunnan 650500, People's Republic of China}

\author[0009-0003-3758-0598]{Liping Li}
\email{liliping@ynao.ac.cn}
\affiliation{Yunnan Observatories (YNAO), Chinese Academy of Sciences (CAS), Kunming, 650216, People's Republic of China}
\affiliation{International Centre of Supernovae, Yunnan Key Laboratory, Kunming 650216, People's Republic of China}
\affiliation{Key Laboratory for the Structure and Evolution of Celestial Objects, CAS, Kunming, 650216, China}

\author[0009-0002-3956-6143]{Qian Zhai}
\email{zhaiqian@ynao.ac.cn}
\affiliation{Yunnan Observatories (YNAO), Chinese Academy of Sciences (CAS), Kunming, 650216, People's Republic of China}
\affiliation{International Centre of Supernovae, Yunnan Key Laboratory, Kunming 650216, People's Republic of China}
\affiliation{Key Laboratory for the Structure and Evolution of Celestial Objects, CAS, Kunming, 650216, China}

\author[0000-0003-1713-0082]{Edoardo P. Lagioia}
\email{elagioia@ynu.edu.cn}
\affiliation{South-Western Institute for Astronomy Research, Yunnan University, Kunming, Yunnan 650504, People's Republic of China}
\affiliation{Yunnan Key Laboratory of Survey Science, Yunnan University, Kunming, Yunnan 650500, People's Republic of China}

\author[0009-0005-5865-0633]{Xueling Du}
\email{xldu@stu.ynu.edu.cn}
\affiliation{South-Western Institute for Astronomy Research, Yunnan University, Kunming, Yunnan 650504, People's Republic of China}
\affiliation{Yunnan Key Laboratory of Survey Science, Yunnan University, Kunming, Yunnan 650500, People's Republic of China}

\author[0000-0002-8700-3671]{Xinzhong Er}
\email{phioen@163.com}
\affiliation{Tianjin Astrophysics Center, Tianjin Normal University, Tianjin, 300387, People's Republic of China}

\author[0000-0001-5258-1466]{Jianhui Lian}
\email{jianhui.lian@ynu.edu.cn}
\affiliation{South-Western Institute for Astronomy Research, Yunnan University, Kunming, Yunnan 650504, People's Republic of China}
\affiliation{Yunnan Key Laboratory of Survey Science, Yunnan University, Kunming, Yunnan 650500, People's Republic of China}

\author[0000-0001-7140-1950]{Ziwei Li}
\email{lzw@mail.ynu.edu.cn}
\affiliation{South-Western Institute for Astronomy Research, Yunnan University, Kunming, Yunnan 650504, People's Republic of China}
\affiliation{Yunnan Key Laboratory of Survey Science, Yunnan University, Kunming, Yunnan 650500, People's Republic of China}

\author[0000-0003-4121-5684]{Shiyan Zhong}
\email{zhongsy@ynu.edu.cn}
\affiliation{South-Western Institute for Astronomy Research, Yunnan University, Kunming, Yunnan 650504, People's Republic of China}
\affiliation{Yunnan Key Laboratory of Survey Science, Yunnan University, Kunming, Yunnan 650500, People's Republic of China}

\author[0000-0003-1295-2909]{Xiaowei Liu}
\email{x.liu@ynu.edu.cn}
\affiliation{South-Western Institute for Astronomy Research, Yunnan University, Kunming, Yunnan 650504, People's Republic of China}
\affiliation{Yunnan Key Laboratory of Survey Science, Yunnan University, Kunming, Yunnan 650500, People's Republic of China}
%------------------------------------------------------------------------------------------------------------------------------


\begin{abstract}
We present the results of low-resolution spectroscopic and densely sampled multi-band photometric follow-up of supernova (SN) 2024aecx. The SN was discovered in the spiral galaxy NGC~3521 (distance $\sim$\,11 Mpc) within a day after the explosion. The early spectra of SN~2024aecx show a weak signature of hydrogen lines, which disappeared in $\sim$30 days after the explosion. Light curves in all bands show a distinct feature of two peaks, and the first peak is likely due to the shock cooling emission. The early phase light curve evolution of SN~2024aecx has similarity with the typical Type IIb events, but the decay rate in different bands (e.g., $\rm \Delta m_{15}$ = 1.60\,$\pm$\,0.05 mag, $g$-band) is significantly faster in the post-peak phase. It attained the secondary maximum in $\sim$19 days ($g$-band) with a peak absolute magnitude of M$_{g}$\,=\,--17.94\,$\pm$\,0.10 mag. 
SN~2024aecx colors trend redder in early epochs ($<$8 days), followed by a duration in which it grows bluer, then later gets redder again $>$20 days after explosion. The analytical model fitting to the light curves reveals an envelope mass and progenitor radii in the range of $\sim$0.03--0.24 $M_\sun$ and $\sim$169--200 $R_\sun$, respectively. Modeling of the pseudo-bolometric light curve suggests that synthesized $^{56}$Ni in the explosion was $\sim$0.15 M$_{\sun}$ with ejecta mass and kinetic energy of $\sim$0.7  M$_{\sun}$ and $\sim$0.16 $\times$ 10$^{51}$ erg, respectively. The observational properties and modeling indicate that the SN~2024aecx progenitor belongs to the extended progenitor category.
\end{abstract}

\section{Introduction}\label{intro}
The evolution of massive stars ($\gtrsim$10 M$_\odot$) ends when their nuclear fuel is exhausted after the formation of iron group elements in the core \citep{Heger2003H, Woosely-1986W}. At this stage, the stellar core collapses \citep{Woosely-2005NatPhW} due to the strong gravitational force on a very short time scale (of the order of a second), and eventually the star explodes as a core-collapse supernova (CC-SN). Various stellar parameters (e.g., mass, metallicity, mass loss, rotation, binary interaction, etc.) may significantly affect the observational features such as light curves and spectra. Accordingly, the observational properties displayed by these transients at early or maximum light phases make a decisive basis for their classification and subclassification \citep{1997filippenko}. A subgroup (Type IIb, Ib, Ic, and Ic-BL) of these events is termed as `stripped-envelope supernovae, SE-SNe' \citep{1941minkowski, Clocchiatti1997C, Galyam2017G, Modjaz2019M}.

In SE-SNe, the hydrogen and/or helium envelopes of their progenitor are partially or completely stripped off before the explosion. In particular, hydrogen lines do not appear in SNe Ib, but helium lines are dominant. SNe Ic exhibit neither of these lines in their spectra, and SNe Ic-BL (broad line) are characterized with broad line velocities ($>$15000 km s$^{-1}$). Type IIb events demonstrate intriguing properties in terms of both spectroscopic and light curve evolution. During the early phases, the hydrogen lines are clearly visible, which disappear progressively in a few weeks \citep{Filippenko1988F}, and the helium lines dominate the spectra at the later phases (similar to SNe Ib).
In addition, double peaks have been observed in the light curves of a handful of IIb SNe, e.g., SN~1993J \citep{1994richmond, Barbon1995B}, SN~2011dh \citep{Arcavi2011A}, SN~2011fu \citep{Kumar2013K}, SN~2013df \citep{2014morales, VanDyk2014V}, SN~2016gkg \citep{Tartaglia2017T, Bersten2018B}, SN~2020bio \citep{Pellegrino2023P}, SN~2022hnt \citep{Farah2025F}, SN~2024iss \citep{2025YamanakaY}, SN~2024uwq \citep{Subrayan2025arXiv250502908S}. While the initial peak is mainly interpreted by the cooling of the extended hydrogen-rich envelope in the post-shock breakout phase \citep[see][]{Falk1977F, Bersten2012B, 2014nakar, Piro2015P, Sapir2017S, Dessart2018-612A..61D}, the secondary peak is mainly powered due to energy released in radioactive decay of $^{56}$Ni $\rightarrow$ $^{56}$Co $\rightarrow$ $^{56}$Fe \citep{Arnett1980A}. However, a few more scenarios were also proposed to explain the SNe IIb first peak: double $^{56}$Ni distribution \citep{Orellana2022}, Thomson scattering and chemical mixing in the SN ejecta \citep{Park2024P}. The double-peak light curve features of SNe IIb have similarities with some `Ca-rich' events which show strong [\ion{Ca}{2}] 7291, 7324 \AA\ emission compared to [\ion{O}{2}] 6300, 6364 \AA\ \citep{Filippenko-2003F, Kawabata-2010, Perets-2010P, Kasliwal-2012K}. The shape of the first peak can infer the radius and mass of the progenitor \citep{2014nakar}. \citet{Chevalier2010C} suggested the extended and compact progenitors of SNe IIb that can produce double-peaked and single-peaked light curves, respectively.  

From the theoretical modeling aspect, Type IIb SNe progenitors can have both single and binary evolution channels. The radiation-driven stellar wind plays a crucial role in stripping the outer envelope \citep{Crowther-2007, Georgy2012G, Yoon-2015AA} in single massive star models, and mass transfer via a binary companion is a preferred path in the binary stellar evolution models \citep{1992podsiadlowski, Nomoto1995N, Claeys2011C, Smith-2014, Sravan2020AS, Dessart2024D}. However, the observational manifestation supports the binary progenitor, which is evident in several direct imaging \citep[e.g.,][and references therein]{Maund2004M, Folatelli2014F, Ryder2018R, VanDyk2014V}.

The prevalence of wide-area surveys in recent years has enabled large, sample-based, qualitative studies of transients, particularly SE-SNe \citep{2015A&A...574A..60T, Liu2016ApJL, 2016MNRAS.457..328L, 2016MNRAS.458.2973P, 2018AA...609A.136T, 2019MNRAS.485.1559P, 2019Shivvers, 2019williamson, Ayala2025A}. Better cadence and sampling are equally useful to corroborate whether there is a transition between some SE-SNe, which were classically divided into different types, and recent studies point toward a continuum between them. Especially in a few cases, SNe Type IIb and Ib indicate a thin boundary \citep[e.g.,][]{Milisavljevic2013M, Gangopadhyay2023G, Dong2024D, Yesmin2025Y}. Therefore, for a better understanding of the observational properties and the progenitor, it is important to study more events in detail, having both common and/or peculiar features.

\begin{figure}
\centering
\includegraphics[width=\columnwidth]{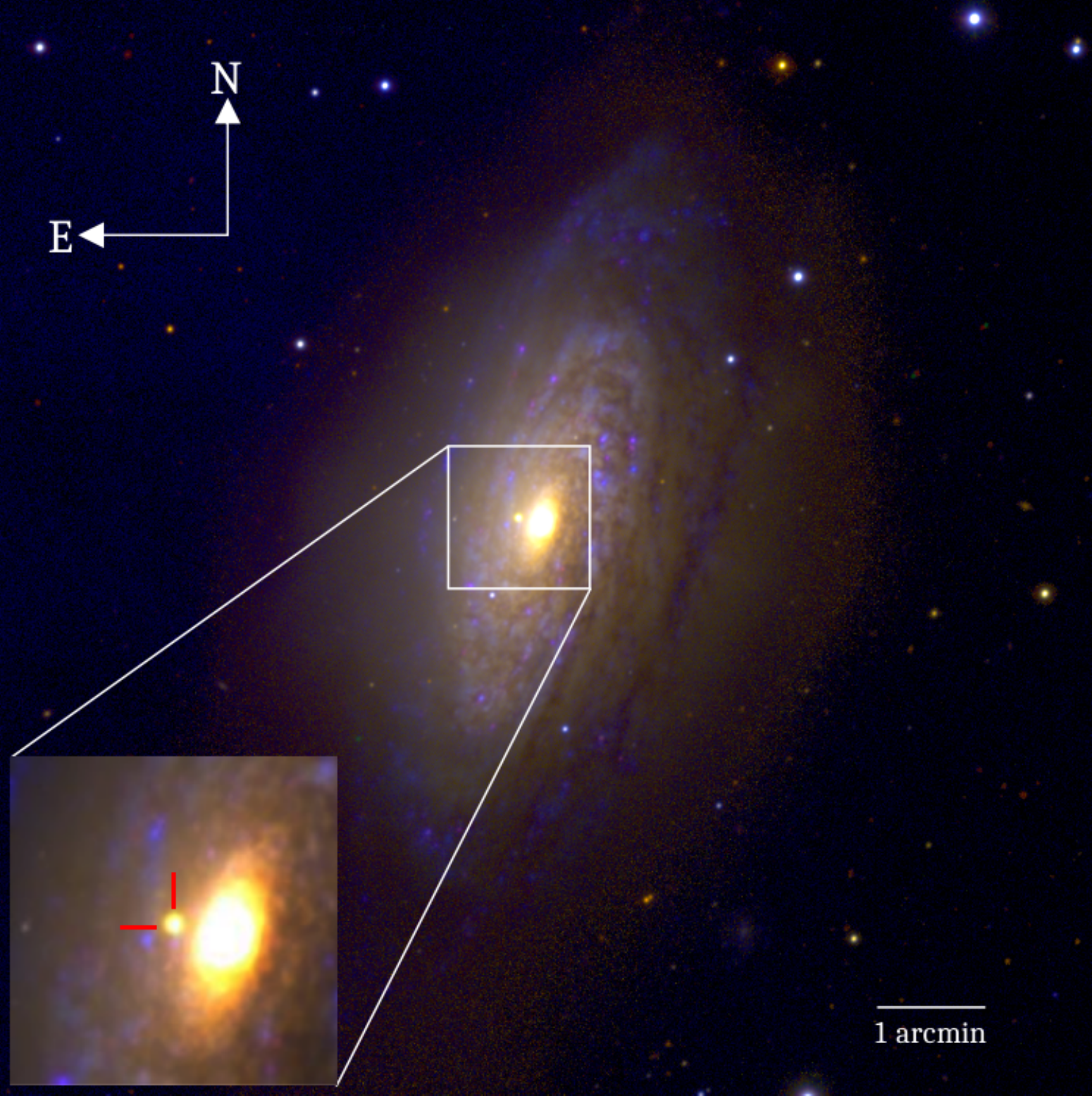}
\caption{SN~2024aecx and the host galaxy NGC~3521. Mephisto's three bands ($u,g,r$) were used to create this RGB image. The zoomed central region of the host is shown in the inset, and SN is indicated with a cross-hair. The displayed image size is $\sim$10\,\arcmin $\times$ 10\arcmin.}
\label{fig:picture}
\end{figure}

In this paper, we investigated the observational properties of a double-peaked Type IIb SN~2024aecx, which exploded in the nearby galaxy NGC~3521 (Fig.~\ref{fig:picture}).

NGC~3521 is a late-type spiral galaxy SAB(rs)bc with a mixed barred and inner ring morphology. \citet[][Third Reference Catalogue of Bright Galaxies, RC3]{deVaucouleurs1991D}, thoroughly examined due to its proximity (11.40 $\pm$ 0.56 Mpc)\footnote{\url{https://ned.ipac.caltech.edu/}}. It has an integrated star formation rate of 2.1 M$_\sun$ year$^{-1}$ \citep{Leroy2008L}, and because of the similarity in structural parameters (stellar mass, rotation velocity, optical radius), it is considered as a near structural analogue to the Milky Way \citep[see][and references therein]{Pilyugin2025P}. The spectroscopic decomposition indicates that NGC~3521 is dominated by the intermediate stellar population \citep{Coccato2018C}. 

SN~2024aecx (ATLAS24rkq) was discovered on MJD~60660.56 (UT 2024 December 16.56) by the ATLAS group in the galaxy NGC~3521 at coordinates (J2000): RA = 11$^{\rm h}$~05$^{\rm m}$~49\fs55 DEC = --\,00\degr ~02\arcmin ~05\farcs44 \citep{Tonry-2024T, Stevance-2024S}. It is one of the very early discovered (within a day) SN after the explosion. The discovery magnitude was 14.68\,$\pm$\,0.01 (orange-ATLAS). The last non-detection was reported on MJD~60659.35 (UT 2024 December 15.35) with a magnitude of 18.56 mag (orange-ATLAS). AT~2024aecx observations (at MJD = 60660.7) from the Lulin Observatory indicated a fading nature and blue colors ($g-r$\,=\,-- 0.6 mag, $r-i$\,=\,-- 0.3 mag) of AT~2024aecx \citep{Chen-2024C-C}. Forced photometry with ATLAS (orange filter) confirmed the fast decline \citep{Fournon-2024P} within a few hours after the explosion.    
We consider the explosion epoch of SN 2024aecx as MJD 60659.953 (2024 December 15.953), the midpoint of the last non-detection, and the discovery date. The considered explosion time matches reasonably with the estimation from the analytical model fitting as discussed in Section~\ref{lc-model}. On 2024 December 17, the SN was classified as a Type IIb based on spectra taken from the SNIFS and GMOS instruments mounted on the 88-inch University of Hawaii telescope \citep{Hinkle-2024H} and Gemini-North telescope \citep{Andrews-2024A1}. However, Gemini-N/GMOS spectroscopy on 2024 December 19 reclassifies it as Type Ic SN \citep{Andrews2024A2}.
Our initial two spectra were acquired on 2024 December 17 (UT) in a gap of $\sim$1 hour (see Section~\ref{sec:obs}). These observations are $\sim$8 hours later than \citet{Andrews-2024A1} and \citet{Hinkle-2024H}. Detailed spectroscopic investigation (Section~\ref{spec-prop}) indicates a weak feature of hydrogen around 6500 \AA, which disappears $\sim$1 month after the explosion. The double-peak multi-band light curves of SN~2024aecx have a close resemblance to Type IIb events (Section~\ref{lc-prop}). Based on these observational characteristics, we consider it a Type IIb SN.

This paper is organized as follows. The observation and data reduction are presented in Section~\ref{sec:obs}, followed by the results in Section~\ref{results}, where photometric and spectroscopic properties are analyzed. The analytical modeling of the early phase light curves and the explosion parameters are presented in Section~\ref{lc-model}, and finally, the results are discussed and summarized in Section~\ref{summary}.

\section{Observations and data reduction}\label{sec:obs}
The photometric and spectroscopic observations of SN~2024aecx are presented in this section. 

\subsection{Photometry}\label{Photometry}

Multi-band photometric observations of SN~2024aecx were started from 2024 December 16 (18:45:01 UT) with 1.6\,m Multi-channel Photometric Survey Telescope \citep[Mephisto,][]{Yuan2020}. The facility is located at the Lijiang Observatory and can perform simultaneous observations of a field in three optical bands at a time ($ugi$ or $vrz$). More details about the Mephisto filters and the instrument can be found in recent studies \citep[][and references therein]{ChenX-2024C, Yuan-PeiY, Yehao2025C, Guowang2025D, Yin2025Y}. Considering a fast evolution in all bands, we acquired images of the SN field on consecutive 3 nights in an interval of 1 to 2 hours. A color composite image of the host, indicating the SN location, is shown in Fig.~\ref{fig:picture}.

Along with 1.6\,m Mephisto, the SN field was also followed up with 50\,cm array facilities (two 50\,cm telescopes, 50-A and 50-B) located near the Mephisto telescope site \citep[see][]{ChenX-2024C}. The 50\,cm array facilities are equipped with an FLI ML 50100 CCD camera of 8176 $\times$ 6132 pixels (size 6 $\mu$m). The filter system is the same in all facilities. The 50-A and 50-B telescopes are equipped with $gr$ and $iz$ filters, respectively, and can simultaneously perform observations in the $gi$ and $rz$ bands.

\begin{table*}
\centering
\small
\caption{Photometric magnitudes of SN~2024aecx$^{\ast}$ (see Section~\ref{Photometry}).}
\label{tab:photometry}
\begin{tabular}{cccccc}
\hline
MJD & Phase$^\dagger$ (days) & Mag & $\sigma_{\mathrm{Mag}}$$^\ast$$^\ast$ & Filter &Telescope \\ \hline
60660.557 & 0.60 & 14.683 & 0.012 & o    & ATLAS \\
60660.559 & 0.61 & 14.761 & 0.011 & o    & ATLAS \\
60660.563 & 0.61 & 14.700 & 0.009 & o    & ATLAS \\
60660.573 & 0.62 & 14.706 & 0.009 & o    & ATLAS \\
60660.769 & 0.82 & 15.175 & 0.035 & z    & 50cm \\
60660.769 & 0.82 & 15.004 & 0.016 & i    & 50cm \\
60660.772 & 0.82 & 15.225 & 0.081 & z    & 50cm \\
60660.777 & 0.82 & 15.214 & 0.022 & i    & 50cm \\
60660.785 & 0.83 & 15.049 & 0.004 & i    & Mephisto \\
60660.785 & 0.83 & 14.666 & 0.002 & g    & Mephisto \\
60660.788 & 0.83 & 14.694 & 0.002 & g    & Mephisto \\
60660.791 & 0.84 & 15.095 & 0.006 & z    & Mephisto \\
60660.792 & 0.84 & 14.914 & 0.002 & v    & Mephisto \\
60660.794 & 0.84 & 15.083 & 0.006 & z    & Mephisto \\
60660.794 & 0.84 & 14.797 & 0.002 & r    & Mephisto \\
60660.795 & 0.84 & 14.825 & 0.002 & v    & Mephisto \\
60660.798 & 0.85 & 15.044 & 0.007 & z    & Mephisto \\
60660.798 & 0.85 & 14.812 & 0.002 & r    & Mephisto \\
60660.798 & 0.85 & 14.917 & 0.002 & v    & Mephisto \\
60660.857 & 0.90 & 15.532 & 0.115 & UVW1 & Swift \\
60660.858 & 0.90 & 14.949 & 0.108 & U    & Swift \\
60660.859 & 0.91 & 15.122 & 0.102 & B    & Swift \\
60660.861 & 0.91 & 16.407 & 0.149 & UVW2 & Swift \\
60660.864 & 0.91 & 15.113 & 0.115 & V    & Swift \\
--       & --   & --    & --    & --  & --        \\
\hline
\end{tabular}\\
$^{\ast}$ This is only a part of the photometric table, and the full table is available as data behind the figure.\\  
$^\dagger$ The phase is relative to explosion epoch (MJD 60659.953).\\
$^\ast$$^\ast$ Mephisto photometric uncertainties primarily comprise photon noise and detector noise.
\end{table*}

The pre-processing of raw frames was performed using a specialized pipeline developed for Mephisto data, which included bias and dark subtraction, flat-fielding, and cosmic ray removal. Afterward, astrometric calibration and photometric measurements were done to extract source information. We used recalibrated Gaia BP/RP (XP) spectra for photometric calibration with the Synthetic Photometry Method \citep{2024ApJS..271...13H, Xiao2023ApJS}. Synthetic AB magnitudes were derived by convolving XP spectra with Mephisto’s transmission curves; the photometric zero-points were determined using high-quality, non-variable stars, correcting for potential gain inaccuracies per CCD output. The calibration uncertainties were better than 0.03 mag in $u$, 0.01 mag in $v$, and 0.005 mag in $griz$ bands. Since SN~2024aecx is deeply embedded inside its host galaxy, we employed image subtraction to remove the influence of the galaxy background on photometry and construct multiband light curves. The specific steps are as follows:

\textit{Image alignment}: As mentioned previously, the NGC~3521 field was observed multiple times with Mephisto before the SN explosion. For Mephisto, we selected good-quality images with smaller Full Width at Half Maximum (FWHM) and low sky background as reference images in each band. The image frames obtained after the SN explosion were considered target images. For the blue ($uv$-bands) and yellow channel ($gr$-bands) images, we extracted $3000 \times 3000~\mathrm{pixel}^2$ ($\sim$\,21.5\arcmin\,$\times$\,21.5\arcmin) cutouts around SN~2024aecx. For the red-channel ($iz$) images, larger $4500 \times 4500~\mathrm{pixel}^2$ ($\sim$\,30\arcmin\,$\times$\,30\arcmin) cutouts were used. For the 50\,cm array telescopes, no pre-explosion images were available, so we adopted high-quality images from 2024 March 27 (UT), when the transient was undetectable, as reference frames. All 50\,cm images were cropped to the same $3000 \times 3000~\mathrm{pixel}^2$ ($\sim$\,21.5\arcmin\,$\times$\,21.5\arcmin) region. Then, science images were aligned relative to the reference images using \texttt{SWARP} \citep{2010ascl.soft10068B}. 

\textit{Image subtraction}: 
To examine our data pipeline, implementation of image subtraction and calibration, etc., we initiated a survey of the nearby galaxies, star clusters, and several pilot fields during the commissioning phase. Consequently, the field of NGC~3521 was observed in multi-bands at various epochs since 2023 January 17. Interestingly, the latest Mephisto observations of NGC~3521 were made on UT 2024 December 09, i.e., about 7 days before the SN discovery (see Section~\ref{lc-prop}).

The image subtraction was performed with the \texttt{HOTPANTS} software \citep{Hotpants-2015B}. For the differencing, we use a convolution kernel with a half-width set to 2 times the image FWHM for the blue ($uv$ band) and yellow ($gr$ band) cameras of Mephisto, and the half-width of the substamp extracted around each centroid is set to 5 times the FWHM. For the 50cm telescope, we adopted the same kernel and substamp scaling (2 and 5 times the FWHM, respectively). For the red channel camera of Mephisto, the kernel half-width is set to 3 times the image FWHM, and the substamp extraction was set to 7 times the FWHM. We set the normalization to the reference image and convolve the reference image to match the target image.

\begin{table*}
\centering
\small
\caption{Spectroscopic observation log of SN~2024aecx.}
\label{tab_spec}
\begin{tabular}{cccccc} \hline
  MJD & Phase$^{\ast}$  (days) & Grism & Spectral range (\AA) & Telescope/Instrument \\ 
\hline
60661.874 & 1.92 & Gr7, Gr8 & 3500--7800, 5200--9250 & HCT/HFOSC \\
60661.924 & 1.97 & Gr3 & 3600--9160 & LJT/YFOSC \\
60664.924 & 4.97 & Gr7, Gr8 & 3500--7800, 5200--9250 & HCT/HFOSC \\
60665.955 & 6.00 & Gr7, Gr8 & 3500--7800, 5200--9250 & HCT/HFOSC \\
60668.854 & 8.90 & Gr7, Gr8 & 3500--7800, 5200--9250 & HCT/HFOSC \\
60670.836 & 10.88 & Gr3 & 3600--9160 & LJT/YFOSC \\
60673.976 & 14.02 & Gr7, Gr8 & 3500--7800, 5200--9250 & HCT/HFOSC \\
60675.818 & 15.87 & Gr7, Gr8 & 3500--7800, 5200--9250 & HCT/HFOSC \\
60681.973 & 22.02 & Gr7, Gr8 & 3500--7800, 5200--9250 & HCT/HFOSC \\
60683.960 & 24.01 & Gr7, Gr8 & 3500--7800, 5200--9250 & HCT/HFOSC \\
60688.972 & 29.02 & Gr7, Gr8 & 3500--7800, 5200--9250 & HCT/HFOSC \\
60689.788 & 29.84 & Gr7, Gr8 & 3500--7800, 5200--9250 & HCT/HFOSC \\
60690.796 & 30.84 & Gr7, Gr8 & 3500--7800, 5200--9250 & HCT/HFOSC \\
60713.909 & 53.96 & Gr7, Gr8 & 3500--7800, 5200--9250 & HCT/HFOSC \\
\hline
\end{tabular} \\
$^{\ast}$ With reference to the date of explosion (MJD 60659.953).
\end{table*}

\begin{table*}
\centering
\small
\caption{Light curve parameters of SN~2024aecx (see Section~\ref{lc-prop}).}
\label{tab_lc_p}
\begin{tabular}{ccccccccc} \hline
Filter & LC min (t$_{m}$)$^{\ast}$ & $\rm m_{min}^{1}$& Decay rate$^{2}$ & LC peak (t$_{p}$)$^{\ast}$ & $\rm m_{max}$ & t$_{rise}^\dagger$  & $\rm \Delta m_{7}$ & $\rm \Delta m_{15}$ \\
       & (day)      &    (mag)         & (mag day$^{-1}$) & (day)   &    (mag)      & (day)        & (mag)            & (mag)              \\  \hline
\textit{Swift U} & 5.91       & 18.36 $\pm$ 0.20 &  0.68 $\pm$ 0.23 & 16.37   & 17.50 $\pm$ 0.01 & 10.46 & --              & --                 \\
\textit{Swift B} & 4.36       & 17.05 $\pm$ 0.08 &  0.56 $\pm$ 0.13 & 17.48   & 15.11 $\pm$ 0.08 & 13.12 & 0.47 $\pm$ 0.15 & --                 \\
\textit{Swift V} & 3.83       & 16.20 $\pm$ 0.04 &  0.37 $\pm$ 0.12 & 19.22   & 14.30 $\pm$ 0.08 & 15.39 & 0.44 $\pm$ 0.12 & --                 \\
$u$              & 7.30       & 19.38 $\pm$ 0.04 &  0.69 $\pm$ 0.04 & 16.72   & 15.57 $\pm$ 0.01 & 9.42  & 0.78 $\pm$ 0.06 & --                 \\
$v$              & 6.35       & 17.68 $\pm$ 0.03 &  0.50 $\pm$ 0.03 & 17.35   & 15.94 $\pm$ 0.01 & 11.00 & 0.86 $\pm$ 0.06 & 2.35 $\pm$ 0.05    \\
$g$              & 4.38       & 16.05 $\pm$ 0.02 &  0.40 $\pm$ 0.02 & 19.34   & 15.55 $\pm$ 0.01 & 14.96 & 0.50 $\pm$ 0.05 & 1.60 $\pm$ 0.05    \\
$r$              & 3.78       & 15.74 $\pm$ 0.02 &  0.32 $\pm$ 0.02 & 20.30   & 15.14 $\pm$ 0.01 & 16.52 & 0.34 $\pm$ 0.04 & 1.16 $\pm$ 0.04    \\
$i$              & 4.15       & 15.63 $\pm$ 0.02 &  0.17 $\pm$ 0.02 & 20.75   & 14.97 $\pm$ 0.01 & 16.60 & 0.20 $\pm$ 0.04 & 1.00 $\pm$ 0.05    \\
$z$              & 3.78       & 15.50 $\pm$ 0.02 &  0.14 $\pm$ 0.02 & 22.56   & 14.97 $\pm$ 0.01 & 18.78 & 0.28 $\pm$ 0.03 & 1.14 $\pm$ 0.04    \\
ATLAS $o$        & 4.10       & 15.72 $\pm$ 0.02 &  0.30 $\pm$ 0.02 & 21.53   & 13.84 $\pm$ 0.01 & 17.43 & 0.46 $\pm$ 0.02 & 1.40 $\pm$ 0.02    \\
\hline
\end{tabular}\\
\begin{minipage}{\textwidth}
$^{\ast}$ Time from the explosion epoch (MJD~60659.953). The maximum possible uncertainty in these estimates is 0.6 days.\\  
$^{1}$ Magnitude at light curve minimum.\\
$^{2}$ Magnitude decay rate between the explosion date and the light curve minimum.\\
$^\dagger$ Rise time from LC min (t$_{m}$) to secondary peak (t$_{p}$), t$_{rise}$\,=\,t$_{p}$--t$_{m}$.\\
\end{minipage}
\end{table*}

\textit{Photometry}: Aperture photometry was performed on the difference images using \texttt{SExtractor} \citep{1996A&AS..117..393B} to construct the light curves. Photometric calibration was carried out using the zero-point values derived from the reference image.

Additionally, we retrieved photometric data ($c, o$ --bands) from the Asteroid Terrestrial-impact Last Alert System \citep[ATLAS,][]{2018PASP..130f4505T, 2020PASP..132h5002S}. For SN~2024aecx, we downloaded these data from the ATLAS forced photometry service\footnote{\url{https://fallingstar-data.com/forcedphot/}}.

SN~2024aecx was also monitored with \textit{Swift}/UVOT \citep{Gehrels2004G,2005roming} in  $uvw2, uvm2, uvw1, u, b$ and $v$ bands. These observations began on 2024 December 16 (UT) and continued up to 2025 January 13 (UT), providing good coverage. We obtained the publicly available data from the Swift Archives\footnote{\url{https://www.swift.ac.uk/swift_portal/}} under the ObsID \textit{00018989}. As the SN was deep inside the host, it required the exclusion of the contribution from the host galaxy. Fortunately, \textit{Swift} had previously monitored the same field under ObsID \textit{00084364}. We downloaded this data and used it for the host templates. We utilized \texttt{Swift\_host\_subtraction}\footnote{\url{https://github.com/gterreran/Swift_host_subtraction}} code \citep{2009brown, 2014brown} to remove the host contribution and perform aperture photometry on the \textit{Swift}/UVOT images.

The photometric magnitudes in different bands (without extinction correction) are listed in Table~\ref{tab:photometry}, and presented in Section~\ref{lc-prop}.

\begin{figure*}
\centering
\includegraphics[scale=0.51]{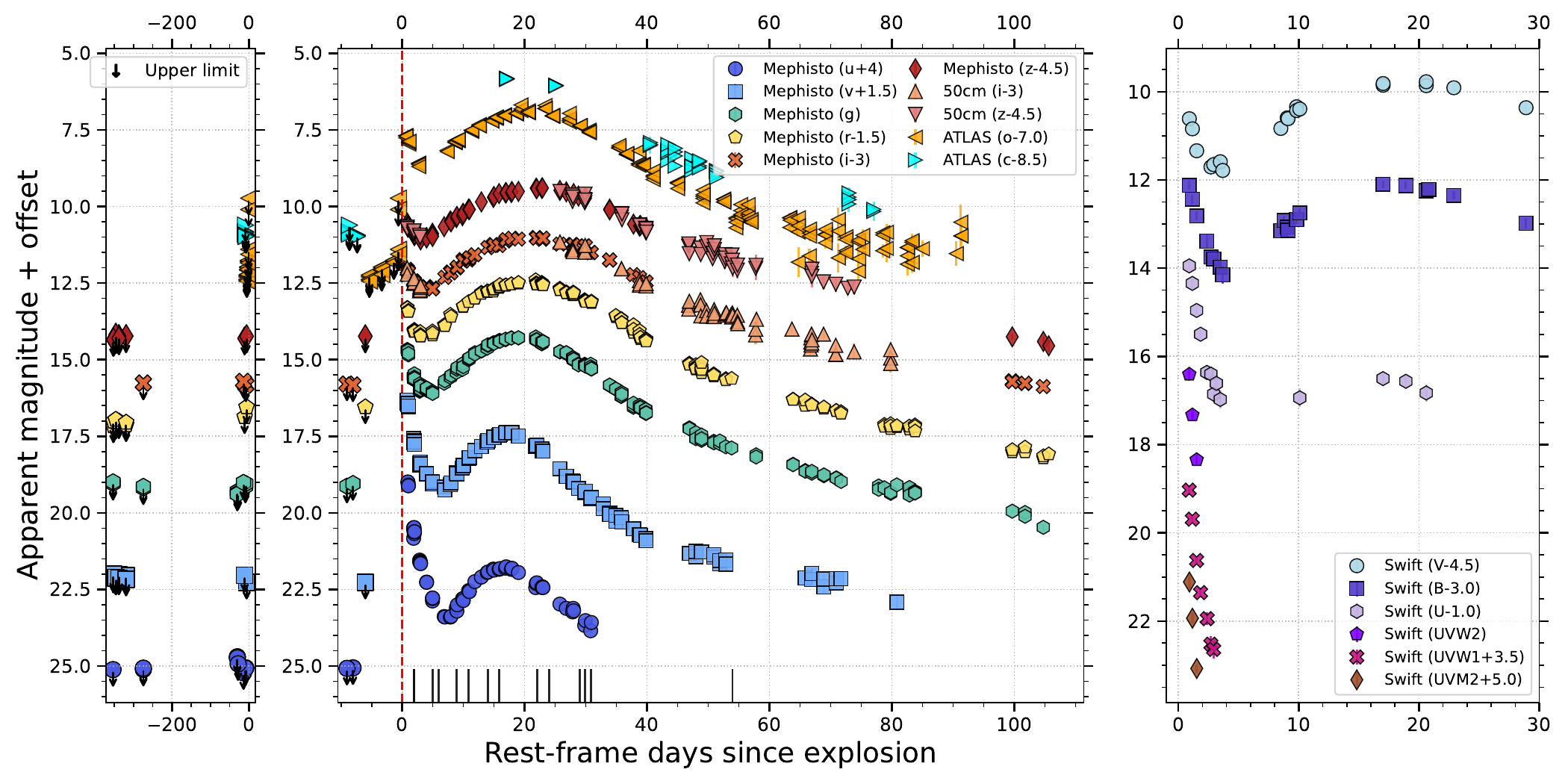}
\caption{Multi-band light curve evolution of SN~2024aecx with respect to the date of explosion. The 3$\sigma$ upper limits (observations taken before the SN explosion, see Section~\ref{obs_sp}) in different bands at the SN location are displayed in the left panel. The template-subtracted SN apparent magnitudes are shown in the middle and right panels. Offsets (indicated in the legend) are applied for clarity. The epoch of explosion is shown with the dashed (red) line. The vertical line segments (black) at the bottom of the middle panel mark the epochs of spectroscopic observations.\\
(The data used to create this figure are available in the online article.)}
\label{fig:fig_lc}
\end{figure*}

\subsection{Spectroscopy}\label{obs_sp}
\subsubsection{2m Himalayan Chandra Telescope (HCT)}
Low-resolution optical spectroscopic observations of SN~2024aecx were initiated with the Himalayan Faint Object Spectrograph Camera (HFOSC) mounted on the $f/9$ Cassegrain focus of 2-meter Himalayan Chandra Telescope (HCT) of the Indian Astronomical Observatory, Hanle, India \citep{2014Prabhu}. Two grisms Gr\#7 (3500--7800 \AA) and Gr\#8 (5200--9250 \AA), available with the HFOSC instrument (spectral resolution, R\,$\sim$600), were used to obtain the spectra. Spectroscopic observations at 12 epochs were obtained between 2024 December 17 (JD~2460662.37) and 2025 February 07 (JD~2460714.41). 

\subsubsection{Lijiang 2.4-meter Telescope (LJT)}

We also obtained two epochs of low-resolution optical spectroscopy with the Yunnan Faint Object Spectrograph and Camera (YFOSC) mounted on the Lijiang 2.4-meter Telescope \citep{Wang-2019RAA}. This facility is located near the Mephisto telescope at Lijiang Observatory and is operated by Yunnan Observatories, Chinese Academy of Sciences (CAS). The grism Gr\#3, which covers the wavelength range of 3400–9100 \AA, and has spectral resolution of $\sim$300, was used to observe SN~2024aecx on 2024 December 17 (JD~2460662.42) and 2024 December 26 (JD~2460671.36).

The spectra of the arc lamp and spectrophotometric standards were obtained, along with the SN for calibration purposes. The spectroscopic data reduction was carried out in a standard way using {\texttt{IRAF}}\footnote{\texttt{IRAF} stands for Image Reduction and Analysis Facility. It is distributed by the National Optical Astronomy Observatory, which is operated by the Association of Universities for Research in Astronomy (AURA) under a cooperative agreement with the National Science Foundation.}. The HCT spectra in both grisms (Gr\#7, Gr\#8) were combined to construct a single flux-calibrated spectrum. The spectra were then scaled with respect to the calibrated $uvgriz$ magnitudes obtained from Mephisto to bring them to an absolute flux scale. Finally, the SN spectra were corrected for the host galaxy redshift of $z$ = 0.002665 $\pm$ 0.000002 (from NED). The log of our spectroscopic observations is provided in Table~\ref{tab_spec}. The data are presented in Section~\ref{spec-prop} and included in the electronic version.

\section{Results}\label{results}

In this section, we present the detailed light curve properties, the extinction in the SN direction, the color and pseudobolometric evolution, and the modeling of the light curves.

\subsection{Evolution of light curves}\label{lc-prop}

Precursor eruptions are common in interacting events such as SNe IIn and Ibn \citep{Foley2007F, Pastorello2007P, Ofek2014O, Reguitti2024R, Bilinski2015B, DongY2024D}. These eruptions may arise months to years before the SN explosion. However, in a sample of 27 SNe IIb from Palomar Transient Factory, only a marginal detection was observed \citep{Strotjohann2015S}. As NGC~3521 was followed up with the Mephisto facility for more than a year, forced photometry was performed at the SN location in multiband images to check any precursor activity before the SN explosion. The limiting magnitudes in different bands are shown in Fig.~\ref{fig:fig_lc} (left panel). We did not notice reliable flux enhancement in different bands at the SN location during the pre-SN monitoring of the host.

The temporal evolution of multi-band light curves is displayed in Fig.~\ref{fig:fig_lc} (middle and right panels). It is evident that there is a decline in all bands just after the explosion, followed by a secondary peak, a typical evolution property of the light curve of SNe IIb \citep[see, for example][]{1994richmond, Kumar2013K, Arcavi2011A, 2014morales, Tartaglia2017T, Pellegrino2023P}. Although the \textit{Swift UVW1, UVM2} band observations are not available after +3 days, a sharp decay can also be seen in those bands (Fig.~\ref{fig:fig_lc}, right panel). Taking into account densely sampled data in different bands, we used the Gaussian Process Regression method to estimate various parameters of the light curve (t$_{m}$, t$_{p}$, m$_{min}$, m$_{max}$, $\rm \Delta m_{7}$, $\rm \Delta m_{15}$) of SN~2024aecx. Here, t$_{m}$ and t$_{p}$ are the time since the explosion of the light curve minimum after the first peak and the secondary peak, respectively. The m$_{min}$ and m$_{max}$ are the magnitudes at t$_{m}$ and t$_{p}$. The parameters $\rm \Delta m_{7}$ and $\rm \Delta m_{15}$ are the decay in magnitudes in 7 and 15 days after the secondary peak. The data points were sometimes extrapolated in the analysis. These parameters are listed in Table~\ref{tab_lc_p}. It is inferred that the duration of the shock cooling phase (t$_{m}$) is longer in the bluer bands than in the redder bands: 7.3 d ($u$-band) and 3.78 d ($z$-band). This is a usual property of double-peaked SNe IIb and has also been seen in SNe~1993J, 2011fu, and 2013df; see Table~4 in \citet{Kumar2013K} and Table~6 in \citet{2014morales}. However, the rise time duration to the secondary peak is opposite, and the bluer bands peaked earlier: 16.72 d ($u$-band) and 22.56 d ($z$-band). Further, it is worth noting that t$_{m}$ (light curve minimum after explosion) of SN~2024aecx is significantly smaller (see Table~\ref{tab_lc_p}) than SNe~1993J, 2011fu \citep[see][]{Kumar2013K, 2015morales} and 2013df \citep{2014morales}. 

The light curve decay rate in post-peak (second) can be determined with $\rm \Delta m_{7}$ and $\rm \Delta m_{15}$ parameters. A faster decline rate signifies less massive ejecta (shorter diffusion timescales). The $\rm \Delta m_{15}$ values for SN~2024aecx are 2.35\,$\pm$\,0.05, 1.6\,$\pm$\,0.5, 1.16\,$\pm$\,0.04, 1.00\,$\pm$\,0.05, 1.14\,$\pm$\,0.04 and 1.40\,$\pm$\,0.02 mag in $vgrizo$ bands, respectively (see Table~\ref{tab_lc_p}). Similarly, the $\rm \Delta m_{7}$ values are also estimated and listed in Table~\ref{tab_lc_p}. In recent sample based studies, $\rm \Delta m_{15}(g)$ was estimated by \citet{Taddia2018T}: 1.10\,$\pm$\,0.10 mag (IIb), 1.25\,$\pm$\,0.19 mag (Ib), and 1.11\,$\pm$\,0.45 mag (Ic). Similarly, \citet{2023ApJ...955...71R} evaluated $\rm \Delta m_{15}(V)$ as 0.88\,$\pm$\,0.16 mag (IIb), 0.81\,$\pm$\,0.24 mag (Ib), and 0.86\,$\pm$\,0.27 mag (Ic). The larger $\rm \Delta m_{15}$ values of SN~2024aecx in different bands indicate that it is one of the fast-evolving events in the post-peak phase than other well-studied SE-SNe.   

\begin{figure*}
\centering
\includegraphics[scale=0.65]{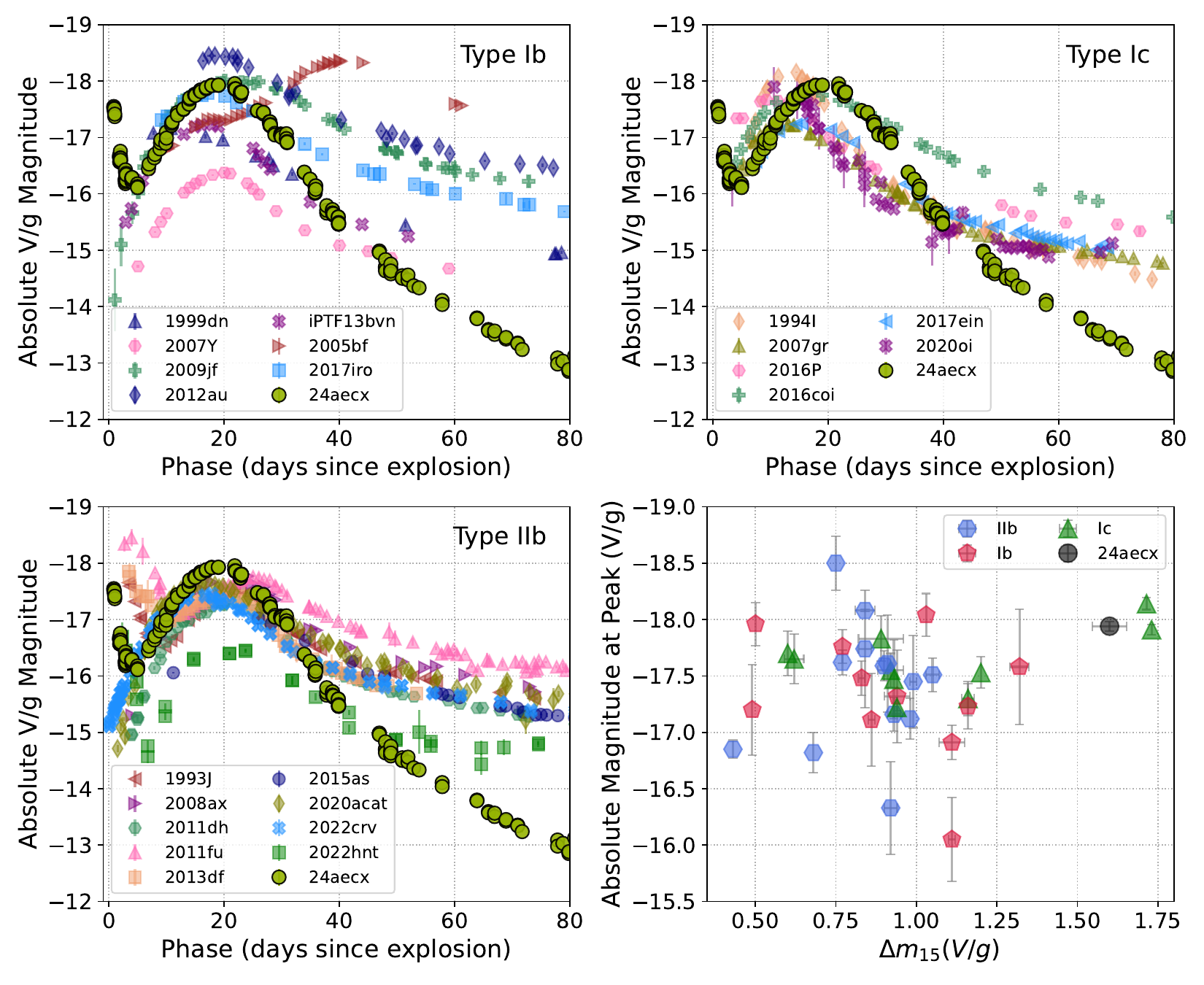}
\caption{The absolute magnitudes ($V$ or $g$ bands) of SNe Ib, Ic, and IIb are compared with SN~2024aecx in the top-left, top-right, and lower-left panels, respectively. The comparison sample is taken from \citet{2018taddia} and the literature (see Section~\ref{lc-prop}). The $V/g$ $\rm \Delta m_{15}$ parameter and absolute $V/g$ band magnitudes are plotted in the bottom-right panel.}\label{fig:fig_abs}
\end{figure*}

In Fig.~\ref{fig:fig_abs}, we have compared the absolute $V$ and $g$ band magnitudes (indicated as $V/g$ in the plot) of Type Ib, Ic, and IIb SNe with SN~2024aecx. The data were obtained from the literature, and the comparison was made with reference to their date of explosion. SNe Type Ib and Ic used in our comparison are SN~1994I \citep{Richmond-1996R}, SN~1999dn \citep{Benetti2011B}, SN~2005bf \citep{Folatelli2006F, Anupama2005A}, SN~2007Y \citep{Stritzinger2009S}, SN~2007gr \citep{Hunter-2009H}, SN~2009jf \citep{Valenti2011V}, SN~2012au \citep{Pandey2021MNRASP}, iPTF13bvn \citep{Srivastav2014S}, SN~2016P \citep{Gangopadhyay-2020G}, SN~2016coi \citep{2018brajesh}, SN~2017ein \citep{Xing-2019X}, SN~2017iro \citep{Kumar2022K}, and SN~2020oi \citep{Rho-2021R}. Similarly, the compared SNe Type IIb are SN~1993J \citep{1994richmond}, SN~2008ax \citep{Pastorello2008P}, SN~2011dh \citep{Arcavi2011A}, SN~2011fu \citep{Kumar2013K, 2015morales}, SN~2013df \citep{2014morales}, SN~2015as \citep{Gangopadhyay2018G}, SN~2020acat \citep{Medler2022MNRASM, Ergon2024A&AE}, SN~2022crv \citep{Gangopadhyay2023G, Dong2024D}, and SN~2022hnt \citep{Farah2025F}.
The estimated absolute $g$ band magnitude for SN~2024aecx at initial epoch and secondary maximum is --17.60\,$\pm$\,0.10 and --17.94\,$\pm$\,0.10 mag, respectively. The comparison plot infers that SN~2024aecx is a luminous SN among the SNe IIb sample and is also comparable to the brighter Ib and Ic events.

In Fig.~\ref{fig:fig_abs} (bottom-right panel), the $\rm \Delta m_{15}$ $V/g$ and absolute $V$/$g$ magnitudes of SE-SNe are plotted. The sample was collected from \citet{2018taddia} and the literature. The distribution is scattered and consistent with other studies \citep{Drout2011D, 2015taddia, Taddia2018T, Kumar2022K, Gangopadhyay2023G}. However, SN~2024aecx occupies the extreme right position along with SN~2020oi and SN~1994I in the plot. During the post-peak phase (20--80 days), SN~2024aecx decline with a rate of $\sim$0.1 mag day$^{-1}$, indicating a faster evolution.

\subsection{Extinction towards SN}\label{reddening}

The reddening due to the Galactic dust along SN~2024aecx line of sight is $E(B-V)$ = 0.05 $\pm$ 0.001 mag \citep*{2011ApJ...737..103S}. Here, the adopted $R_{V}$ is 3.1 following the extinction law of \citet{Cardelli}. The fact that SN is located deep inside the host galaxy, strong reddening is expected. Owing to the absence of any reliable technique for estimation, it is challenging to determine the extinction in such a scenario accurately. 
To constrain the host galaxy extinction, we use the equivalent width (EW) of the Na\,{\sc i} D lines as a proxy for reddening \citep{2003fthp.conf..200T, Stritzinger2018S, 2023ApJ...955...71R}.
We estimate the Na\,{\sc i} D line EW 2.7 $\pm$ 0.5 \AA\ in the spectra near maximum light ($\sim$--2 d) and applied the relation from \citet{Stritzinger2018S}: $A_{V}$(host) = 0.78 ($\pm$0.15) $\times$ EW$_{Na\,{I}}$ D [\AA] and thus $E(B-V)$ = 0.68 $\pm$ 0.22 mag for the host (for $R_{V}$ = 3.1). To further constrain the host galaxy extinction, we attempted to match the near-maximum light spectra of iPTF13bvn \citep{Cao-2013C, Srivastav2014S} with our similar epoch spectra. SN~2024aecx spectra match quite well for $E(B-V)$ value between 0.35 and 0.45 mag (total extinction). Additionally, for a total $E(B-V)$ = 0.45 mag, the $B-V$, $V-R$ color evolution of SN~2024aecx matches reasonably well with the average color evolution of SE-SNe templates from the Carnegie Supernova Project compiled by \citet{Stritzinger2018S}. Therefore, we consider a total (host and Milky Way) $E(B-V)$ = 0.45 mag in our analysis, whereas $E(B-V)$ = 0.68 mag as an upper limit.

\begin{figure}
\includegraphics[width=8.5cm]{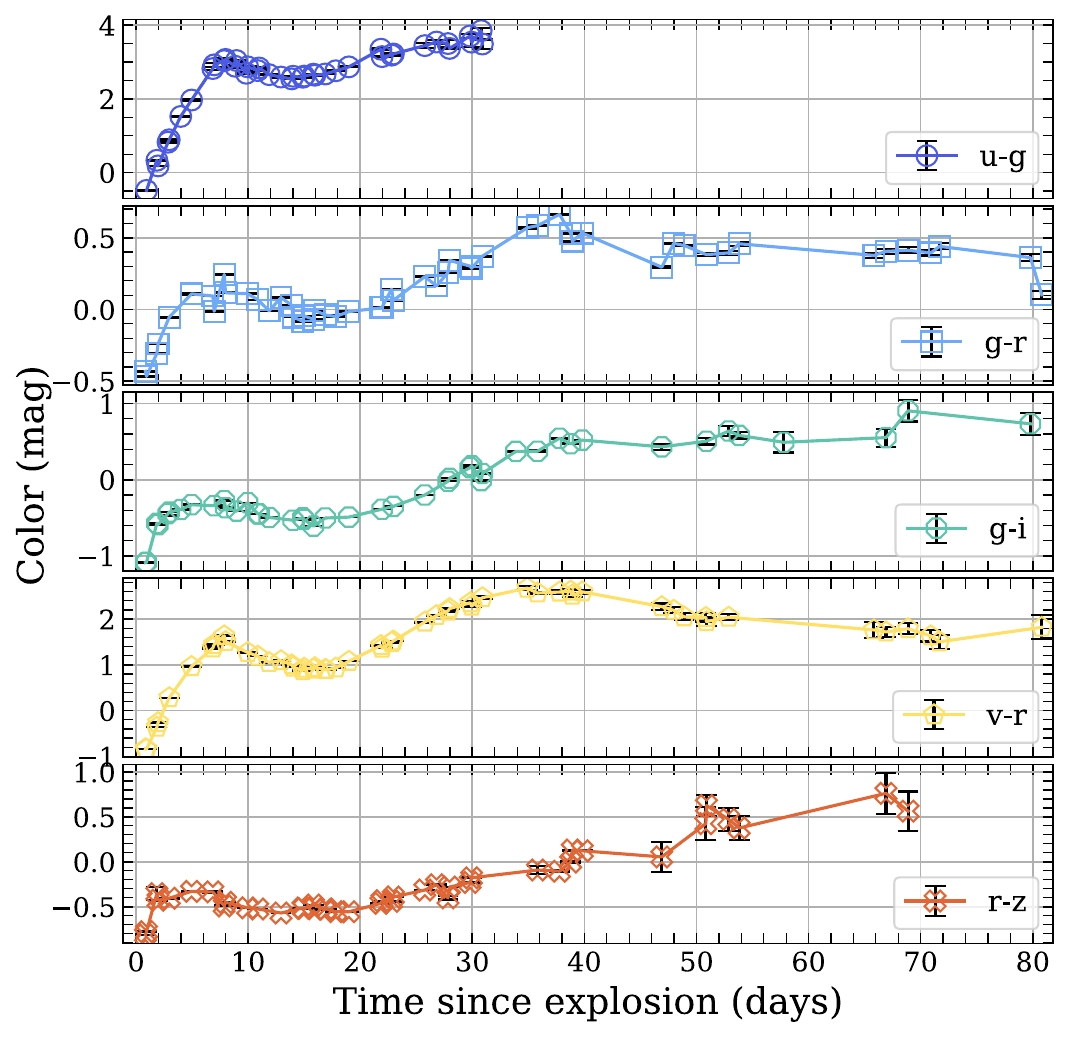}
\caption{The color evolution ($u-g, g-r, g-i, v-r$ and $r-z$) of SN~2024aecx is displayed with different symbols. The colors are corrected for the total reddening $E(B-V )$ = 0.45 mag (see, Section~\ref{reddening}).}
\label{fig:colorcurve}
\end{figure}

\begin{figure}
\includegraphics[width=8.5cm]{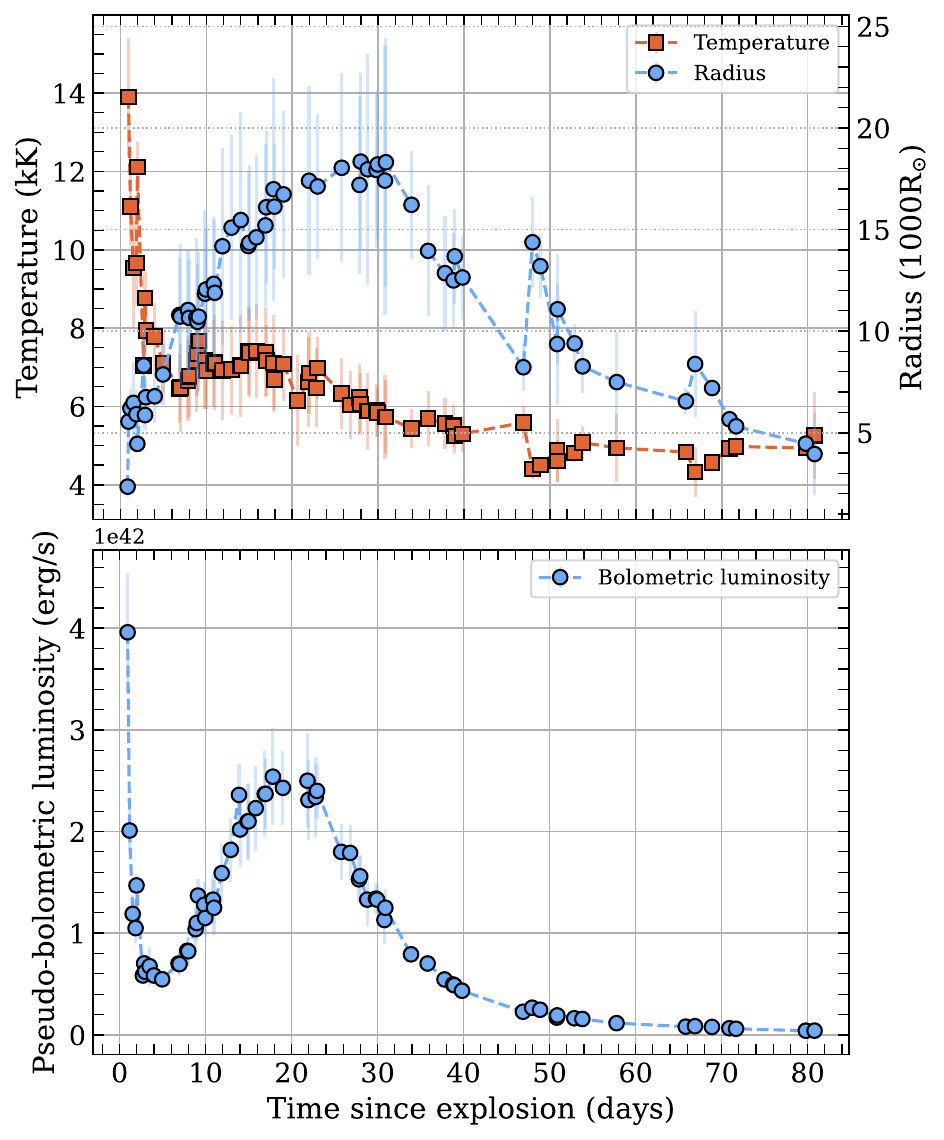}
\caption{The BB temperature and radius evolution of SN~2024aecx (top panel), and the pseudobolometric luminosity (bottom panel).}
\label{fig:temp-radius}
\end{figure}

\subsection{Color indices and Pseudo-bolometric light curve}\label{lc-Bol}

The temperature evolution of SN can be very fast at early times \citep[c.f. adiabatic cooling dominating the LC][]{Woosley1994W} and thus influence different colors. Good cadence data obtained from Mephisto observations were used to estimate the color evolution of SN~2024aecx. The intrinsic colors (corrected for total extinction) $u-g, g-r, g-i, v-r$ and $r-z$ are shown in Fig.~\ref{fig:colorcurve}. The color evolution can be separated into four phases: $\sim$0--8, $\sim$8--18, $\sim$18--40, and beyond $\sim$40 days after the explosion. It is evident that all colors indicate an evolution trend of blue-red up to $\sim$8 days. Interestingly, SNe 1993J, 2011dh, 2013df, 2016gkg and 2024uwq have also shown a similar color evolution in the bluer bands ($B-V$) at initial epochs \citep{Kumar2013K, 2014morales, 2015morales}, although for a shorter duration, $\sim$5 days \citep[see][]{Tartaglia2017T}. 

The red-blue evolution pattern of SN~2024aecx noticed between days $\sim$8 and 18 is rare and only observed in SN~2016gkg and SN~2024uwq. There is a reversal of color ($g-r, g-i, v-r$) evolution between $\sim$18--40 days. Later (after $\sim$40 days), the colors $g-r$ and $v-r$ became redder, but no significant variation was observed in the $g-i$ color in that phase. The $r-z$ color evolution of SN~2024aecx shows gradual reddening between days $\sim$20 to $\sim$70 and then follows a blueward trend. It is noted that in our color estimates, the accuracy is $\sim$3$\%$ after applying the propagation of the error of the photometric measurements. \citet{Subrayan2025arXiv250502908S} have recently investigated the observational properties of SN~2024uwq (IIb). There is a good resemblance in the evolution of different colors between SN~2024aecx and SN~2024uwq (see their Figure~3).

To construct the pseudobolometric light curve of SN~2024aecx, we used the bolometric module of the open-source package \texttt{lightcurve-fitting} \citep{hosseinzadeh_2024_11405219}. Here, we emphasize that although shorter wavelength magnitudes are available only for a limited duration but ultraviolet radiation significantly contributes to the light curves in the early phases. Therefore, all available magnitudes, including $Swift$, were utilized for pseudobolometric light curve estimation. The total extinction towards SN~2024aecx and adopted distance (11.40 Mpc) were used in pseudobolometric calculations (see Section~\ref{reddening}). 
Here, it is worth mentioning that the bolometric luminosity in SE-SNe is mainly dominated by optical emission, but the UV flux may contribute significantly in early phases. For example, in SN~2024iss the UV emission contributed up to $\sim$40$\%$ between 0\,--\,7 days post explosion and decreased rapidly below $\sim$10$\%$ in a few weeks \citep{2025YamanakaY}. Similarly, the NIR emission fraction also varies from $\sim$20$\%$ (early phases) to $\sim$35--40$\%$ ($\sim$1 month after explosion) \citep{Medler2022M, 2025YamanakaY}. However, for estimating the pseudobolometric luminosity of SN~2024aecx, we only use the UV and optical band fluxes, and the NIR contribution was not included due to the lack of observations in these bands. Therefore, we caution the reader to consider it as a lower limit only.

\begin{figure*}
\centering
\includegraphics[scale=0.53]{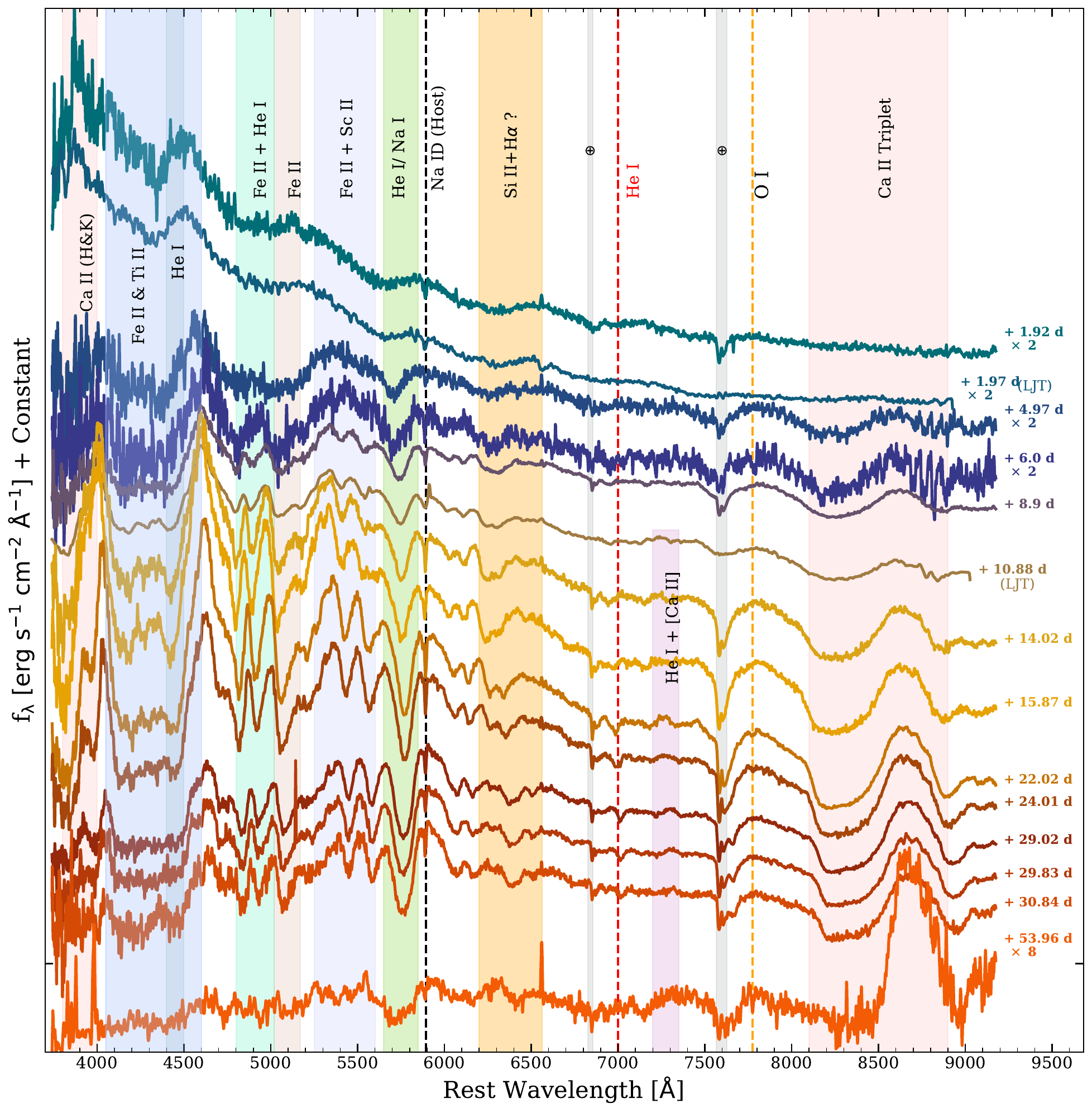}
\caption{The spectroscopic evolution of SN~2024aecx. Prominent spectral lines are marked. The indicated phases (on the right-hand side) are with respect to the date of explosion (see Section~\ref{intro}). Major telluric bands are shown with a circled plus sign. (The data used to create this figure are available in the online article.)}
\label{fig:fig_spec}
\end{figure*}

The result is plotted in Fig.~\ref{fig:temp-radius} (bottom panel). The initial point of the shock cooling phase has log$_{10}L$\,=\,42.60\,$\pm$\,0.03 erg s$^{-1}$ and it decreased to a minimum of log$_{10}L$\,=\,41.65\,$\pm$\,0.02 erg s$^{-1}$ within $\sim$5 days after explosion. The second peak was reached afterward on MJD 60679.7\,$\pm$\,0.3 with log$_{10}L$\,=\,42.39\,$\pm$\,0.02 erg s$^{-1}$. The blackbody (BB) fitted temperature and radius parameters are also shown in Fig.~\ref{fig:temp-radius} (top panel). This indicates an abrupt decline in the temperature from $\sim$14 kK to 6.5 kK in $\sim$8 days, which is consistent with blue colors shown in Fig.~\ref{fig:colorcurve} in the same duration. Then it gradually reached the secondary maximum ($\sim$7 kK) and finally settled around 5 kK in the later phases. The shock cooling phase BB temperature evolution of SN~2024aecx resembles nicely with SN~2016gkg \citep{Tartaglia2017T}. There is a gradual change in BB radius, which attains a maximum $\sim$12000\,$\times\,R_{\sun}$ around day 30 and slowly declines at later epochs.

\begin{figure}
\centering
\includegraphics[scale = 0.48]{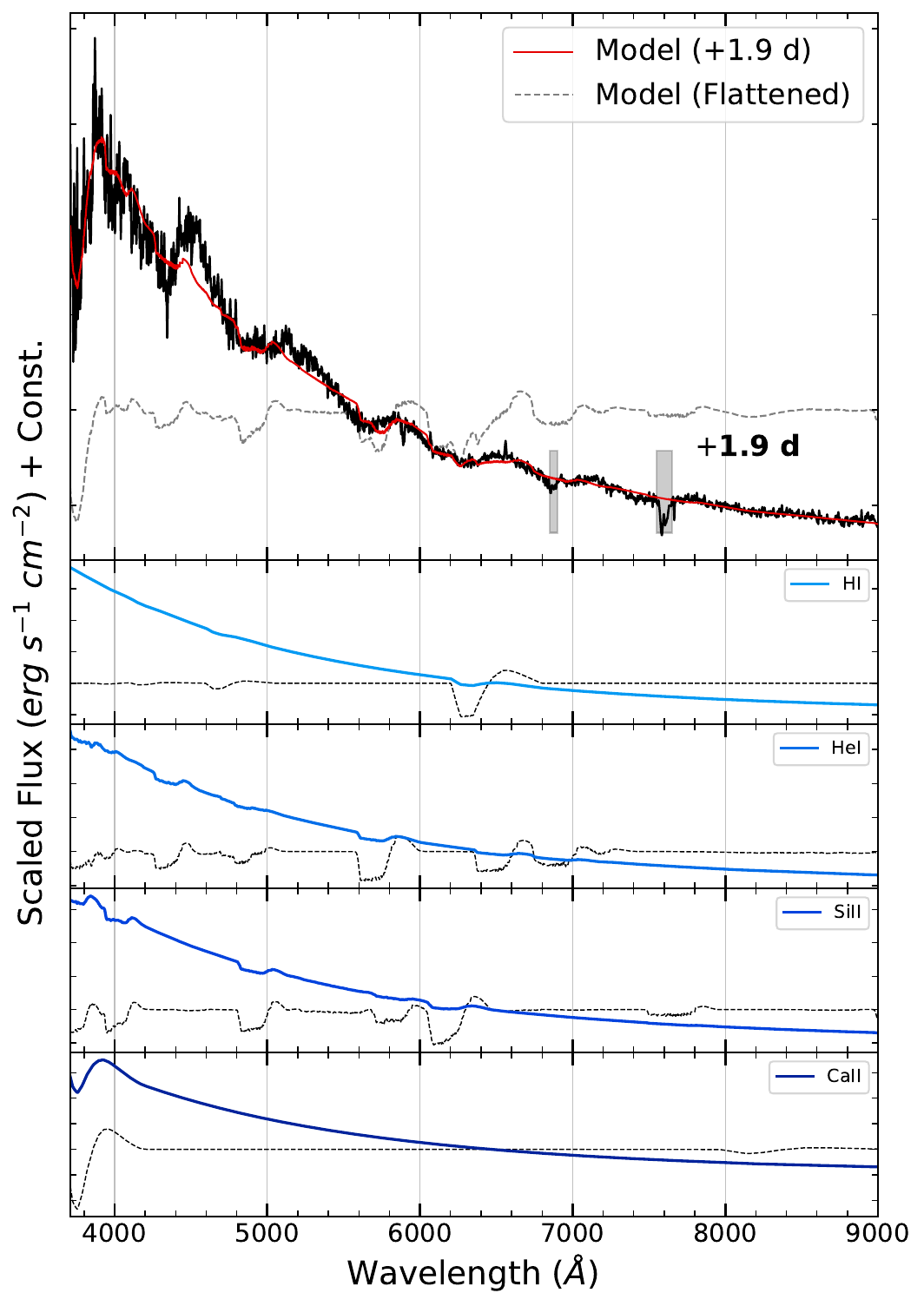}
\caption{\texttt{SYNAPPS} fit of the +1.9~d spectrum of SN~2024aecx. The \texttt{SYNAPPS} fit (red) is overplotted to the real data (black) in the top panel. Subsequent panels show the contributions from the individual elements. The plot also shows the flattened spectra for clarity. (The gray shaded regions were masked during the fitting.)}
\label{synapps-fit}
\end{figure}

\begin{figure}
\centering
\includegraphics[scale = 0.39]{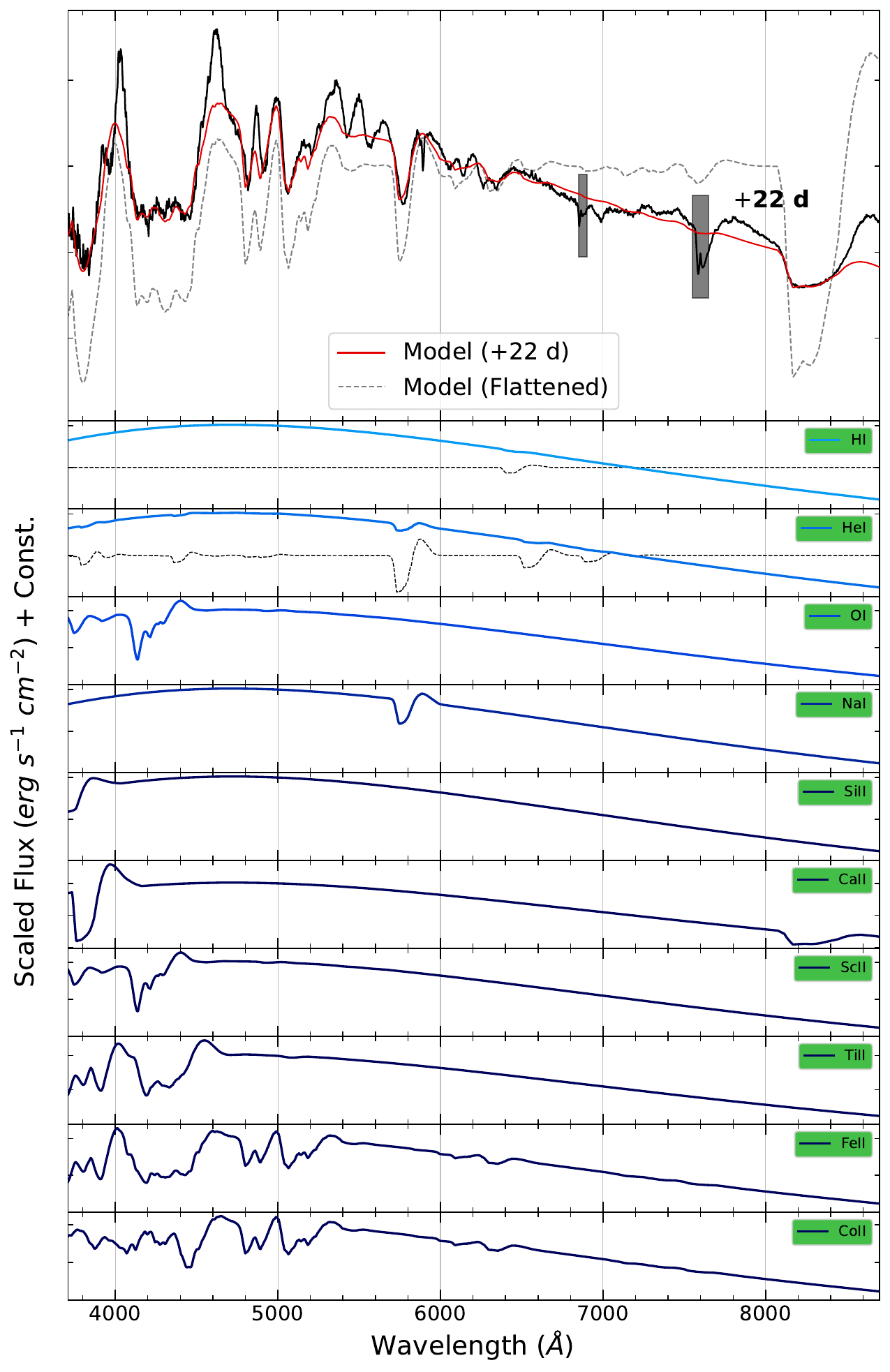}
\caption{Similar to Fig.~\ref{synapps-fit} but the \texttt{SYNAPPS} fit is for +22 d spectrum of SN~2024aecx. Additional elements were added here as indicated in the figure.}
\label{synapps-fit-2}
\end{figure}

\subsection{Spectral evolution}\label{spec-prop}
We present the complete spectroscopic evolution of SN~2024aecx in Figure~\ref{fig:fig_spec}. All the spectra have been corrected for host redshift, calibrated with SN photometry, and dereddened for total extinction estimated in Section~\ref{reddening}. The spectral sequence spans a time period of around 55 days with respect to the explosion epoch. It covers both the shock cooling and secondary peak phases. The initial two spectra at +~1.9 and +~2.0~days show a very blue continuum with some underdeveloped broad features, which are possibly due to hydrogen/silicon and helium. We confirm the presence of H$\alpha$/\ion{Si}{2} (6563 \AA, 6355 \AA) and He (5876 \AA) in these early spectra using \texttt{SYNAPPS} modeling, which we discuss later. Below 4000~\AA, we observe a prominent emission attributed to \ion{Ca}{2} H\&K (3934 and 3968 \AA). As the ejecta evolves further, we see rapid cooling, making the continuum redder. 

Several metallic lines start to form with strong absorptions around +5 day. However, our spectra around +5-6~days have slightly poorer SNR than the rest of the spectra owing to the decline in brightness during the shock cooling minima. We find conspicuous features from \ion{Fe}{2} and \ion{Sc}{2} towards the bluer region. Around 8000--9000~\AA, we also observe the development of \ion{Ca}{2} triplet. The Na\,{\sc i} D line arising due to the host is prominently visible and used to constrain the extinction towards SN (see Section~\ref{reddening}). Several prominent features identified are marked in the spectra sequence Fig.~\ref{fig:fig_spec}.

Around +~16~day, most of the lines are fully developed with bluer side spectra dominated by several \ion{Fe}{2} lines, whereas the redder side region is comparatively devoid of lines except the broad absorption and emission feature due to \ion{Ca}{2} triplet. Even the earlier broad emission feature around 6500~\AA\, shifts to a narrow absorption feature. We further observe a broad absorption band which could be due to multiple lines from \ion{Fe}{2} and \ion{Ti}{2}. 

\subsection{\texttt{SYNAPPS} models}\label{Synapps}

\begin{figure*}
\centering
\includegraphics[scale = 0.41]{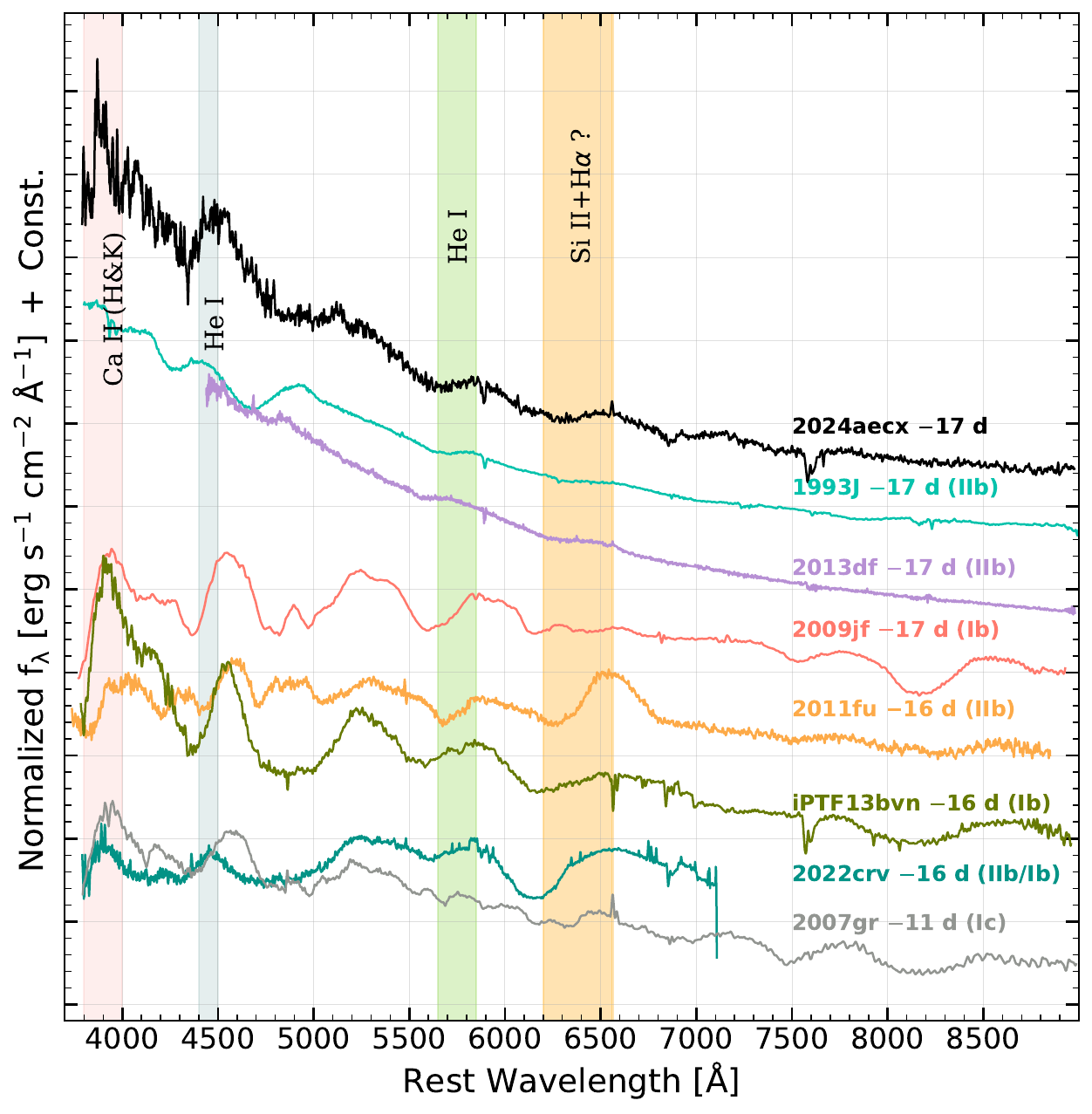}
\includegraphics[scale = 0.41]{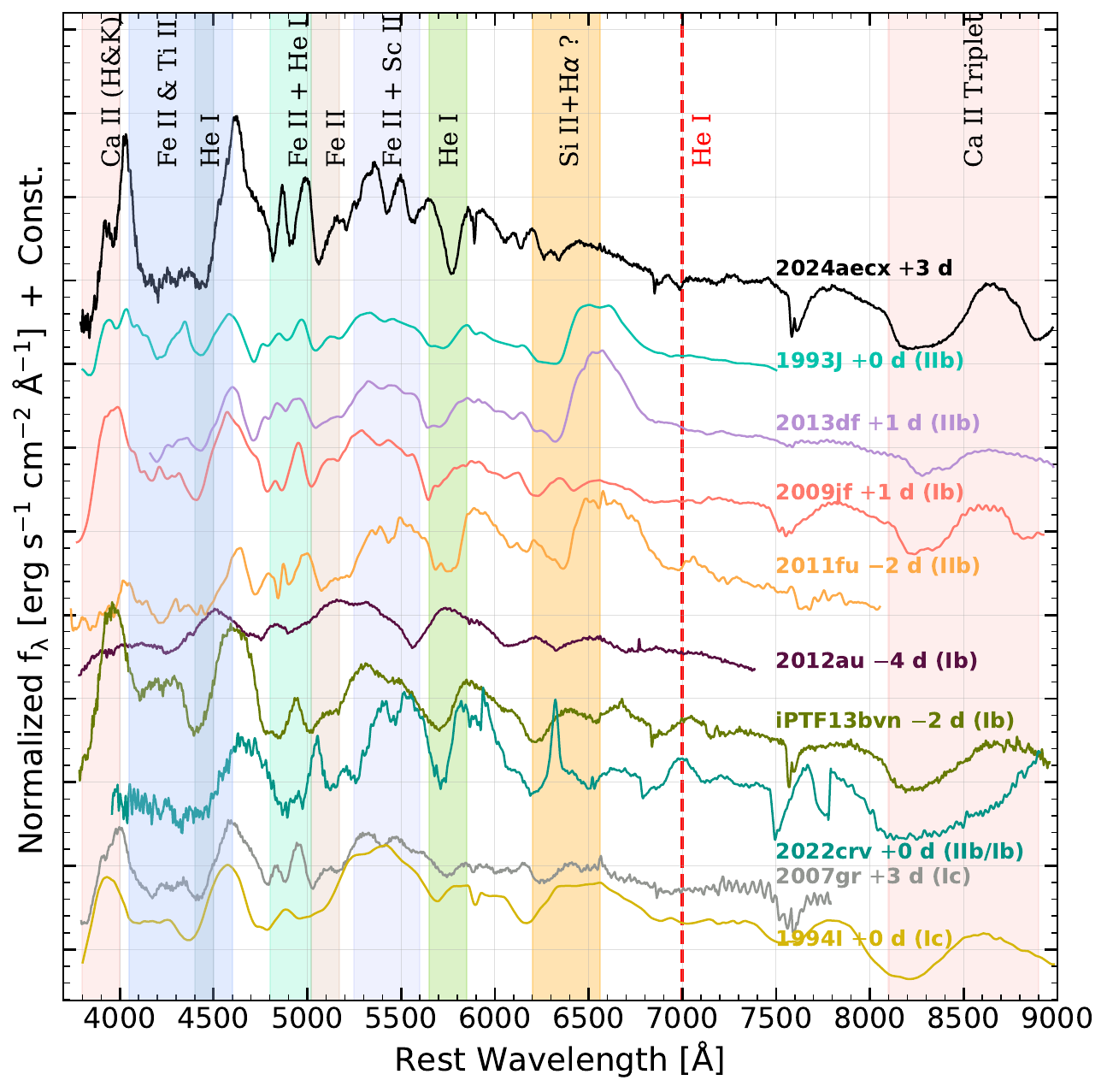}
\caption{Comparison of SN~2024aecx spectra with Type Ib, Ic, and IIb events (SNe~1993J, 1994I, 2007gr, 2009jf, 2012au, 2013df, iPTF13bvn, 2022crv) at similar epochs. The indicated phases are with respect to the maximum epoch, and prominent spectral lines are marked.}
\label{spec-comp}
\end{figure*}

Owing to uncertainty and ambiguity in some of the identified lines, we attempt to model the early spectrum (+\,1.9 day) and the spectrum around the maximum (+\,22 day) to constrain some of the spectral features. For this purpose, we make use of the open-source \texttt{SYNAPPS} code \citep{2011PASP..123..237T}. This is a direct implementation of another code called \texttt{SYNOW} \citep{2010ascl.soft10055P}, which is a parametrized spectral synthesis code.  
A spherical symmetry with ejecta being in homologous expansion is assumed in this setup. The photons are emitted from a sharp photosphere, with the optical depth being described as an exponential function of velocity.
$ \tau_{ref}(v)=\tau_{ref}(v_{ref})\exp{\left (\frac{v_{ref}-v}{v_e} \right )}$ where $v_{ref}$ is reference velocity for parameterization and $v_e$ is the maximum velocity allowed at the outer edge of the line-forming region \citep{2011PASP..123..237T}. For some optical depth, the reference line profile is estimated for a given ion with the remaining lines following Boltzmann statistics \citep{2010ascl.soft10055P}. \texttt{SYNAPPS} automatically generates synthetic spectra based on a provided set of inputs, including parameters such as ions, blackbody temperature, velocities, and opacities. It continues to iterate over these parameters within pre-defined ranges and compares them with the observed spectra for each iteration to obtain a best-fit model spectrum.

For the first spectrum at +1.9 d, we use the following elemental species \ion{H}{1}, \ion{He}{1}, \ion{Na}{1}, \ion{Si}{2}, \ion{Ca}{2}, and \ion{Fe}{2}. The best describing model spectrum, along with the contribution from the individual species, is shown in Figure~\ref{synapps-fit}. We also show a flattened spectrum for the model and individual species to better contrast the contributions. As expected, we did not find any contribution from the Na and Fe lines for the broad features at such an early phase, hence not shown in sub-panels as well. For this model, the lines are best fit for a photospheric velocity of $\rm \sim14,000~km~s^{-1}$ and a temperature of $\rm \sim 14600~K$.

We also model the SN~2024aecx spectrum around maximum ($\rm +22~day$), where we find the maximum number of elements with well-defined line profiles. The best describing model fits are presented in Figure~\ref{synapps-fit-2}. As is evident, many of the spectral features are satisfactorily explained with the model fits. The photospheric velocity and temperature obtained at this phase are  $\rm \sim7700\,km~s^{-1}$ and $\rm\sim6200~K$, respectively. The region blueward of 5500~\AA~ is mostly dominated by multiple blended features from various metals, specifically \ion{Fe}{2}, \ion{Ti}{2}, and  \ion{Co}{2} with some contributions from \ion{Sc}{2} and \ion{Ca}{2} as well. The strong absorption around 5900~\AA~ is well defined by \ion{Na}{1} line along with some contribution from \ion{He}{1}. There are some other features arising due to \ion{He}{1} as seen in the flattened model spectra for He, but due to significant blending from other metal features, it is not trivial to discern the He features in those regions. Similarly, there is some hint of H in the spectra during peak, as we can clearly see the H feature around ~6500~\AA\, in the model fits. However, we could not trace any other Balmer feature (e.g., H$\beta$ 4861 \AA) apart from the weak H$\alpha$ feature. 

In Fig.~\ref{spec-comp}, we compare SN~2024aecx spectra at similar epochs with other well-studied SNe Ib, Ic, and IIb indicated in the figure. The comparison spectra (SNe~1993J, 1994I, 2007gr, 2009jf, 2013df, 2012au, iPTF13bvn, 2022crv) were retrieved from the respective studies (see Section~\ref{lc-prop} for references) and from WISeRep \citep{2012yaron}. The earliest spectrum (+1.9~d) of SN~2024aecx, which belongs to the shock cooling phase, has similarity with SN~1993J and SN~2013df (see left panel of Fig.~\ref{spec-comp}). These are blue and almost featureless. Here, the blue feature is also evident from the early colors as discussed in Section~\ref{lc-Bol}. However, at a similar epoch, line formation has already begun in SNe 2007gr, 2009jf, 2011fu, iPTF13bvn, and 2022crv. The spectra near the secondary maximum are plotted in the right panel of Fig.~\ref{spec-comp}. By this time, the various spectral lines, such as \ion{Fe}{2}, \ion{He}{1}, and \ion{Sc}{2} can be clearly identified in all events. Notably, the \ion{Fe}{2} line near 5000 \AA\, is more prominent in SN~2024aecx than in other events.

\subsubsection{Line velocities}

\begin{figure}
\centering
\includegraphics[width=\columnwidth]{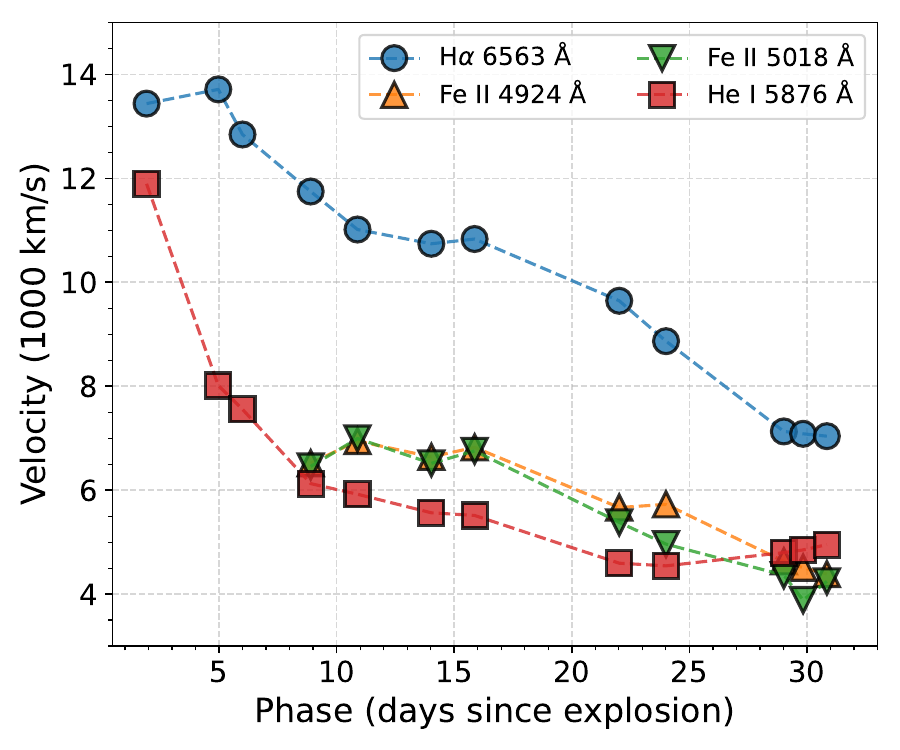}
\caption{Line (H$\alpha$, \ion{He}{1} and \ion{Fe}{2}) velocity evolution of SN~2024aecx are plotted with different symbols (see Section~\ref{spec-prop}). The measurement errors are not shown here, but they can be up to 600 km s$^{-1}$.}
\label{Line-vel}
\end{figure}

Major spectral line velocities of SN~2024aecx such as H$\alpha$ (6563 \AA), \ion{He}{1} (5876 \AA), \ion{Fe}{2} (4924 \AA, 5018 \AA) and \ion{Ca}{2} (8498 \AA) were estimated by Gaussian profile fitting procedure to the absorption trough of these lines. The spectra were corrected for the host redshift before the measurements. The evolution of line velocities of different elements is presented in Fig.~\ref{Line-vel}. The H$\alpha$ velocity was $\sim$14000 km s$^{-1}$ at the early phases, then it gradually decreased $\sim$10000 km s$^{-1}$ near the secondary peak and finally settles around 7000 km s$^{-1}$ (day $\sim$30). It is important to note that during the shock cooling phase, the H$\alpha$ velocity of SNe~1993J, 2011fu was significantly higher than SN~2024aecx, but it is close to SNe~2011dh, 2013df, and 2016gkg. 

The \ion{He}{1} velocity evolved from 12000 to $\sim$5000 km s$^{-1}$ between days 1 to 30 and was always below H$\alpha$ velocities. There is a similarity in \ion{He}{1} velocity values of SN~2024aecx to other IIb SNe \citep{Liu2016ApJL}. Interestingly, in SN~2024aecx the \ion{He}{1} line evolved from the beginning; nevertheless, in SN~2013df it started appearing $\sim$10 days after the explosion \citep{2014morales}. The velocity values of both \ion{Fe}{2} (4924 \AA, 5018 \AA) lines have slower evolution and lie between 7000 to 5000 km s$^{-1}$. The \ion{Ca}{2} 8498 \AA\, have a velocity of $\sim$10000 km s$^{-1}$ around day 30. Both line velocities of \ion{Fe}{2} (4924 \AA) and \ion{Fe}{2} (5018 \AA) are consistent. Notably, there is no definite line that can be treated directly for the photosphere; however, \ion{Fe}{2} traces it \citep{2005bdessart}. The average of \ion{Fe}{2} velocities between day 16 and 22 was considered as the photospheric velocity near maximum and used to calculate the kinetic energy of the explosion in Section~\ref{lc-model}.  

\begin{table*}
\centering
\caption{Results of shock cooling model fitting (see Section~\ref{lc-model}).}
\label{LC-fitting}
\begin{tabular}{ccccc} \hline
Polytropic index (n) & Radius (R$_\sun$)  & M$_{en}$ (M$_\sun$)   &  v$_{sh}$ (10$^{9}$ cm s$^{-1}$)  &  t$_{0}$ (day) \\ \hline
3                    & 169.15$^{+1.23}_{-1.27}$ & 0.24$^{+0.0003}_{-0.0004}$   &  2.35$^{+0.003}_{-0.003}$         & --0.55$^{+0.005}_{-0.005}$   \\
3/2                  & 200.35$^{+0.73}_{-0.70}$   & 0.03$^{+0.00003}_{-0.00003}$ &  1.55$^{+0.002}_{-0.002}$         & --1.00$^{+0.0002}_{-0.0001}$ \\
\hline
\end{tabular}\\
\end{table*}

\begin{figure*}
\centering
\includegraphics[scale = 0.5]{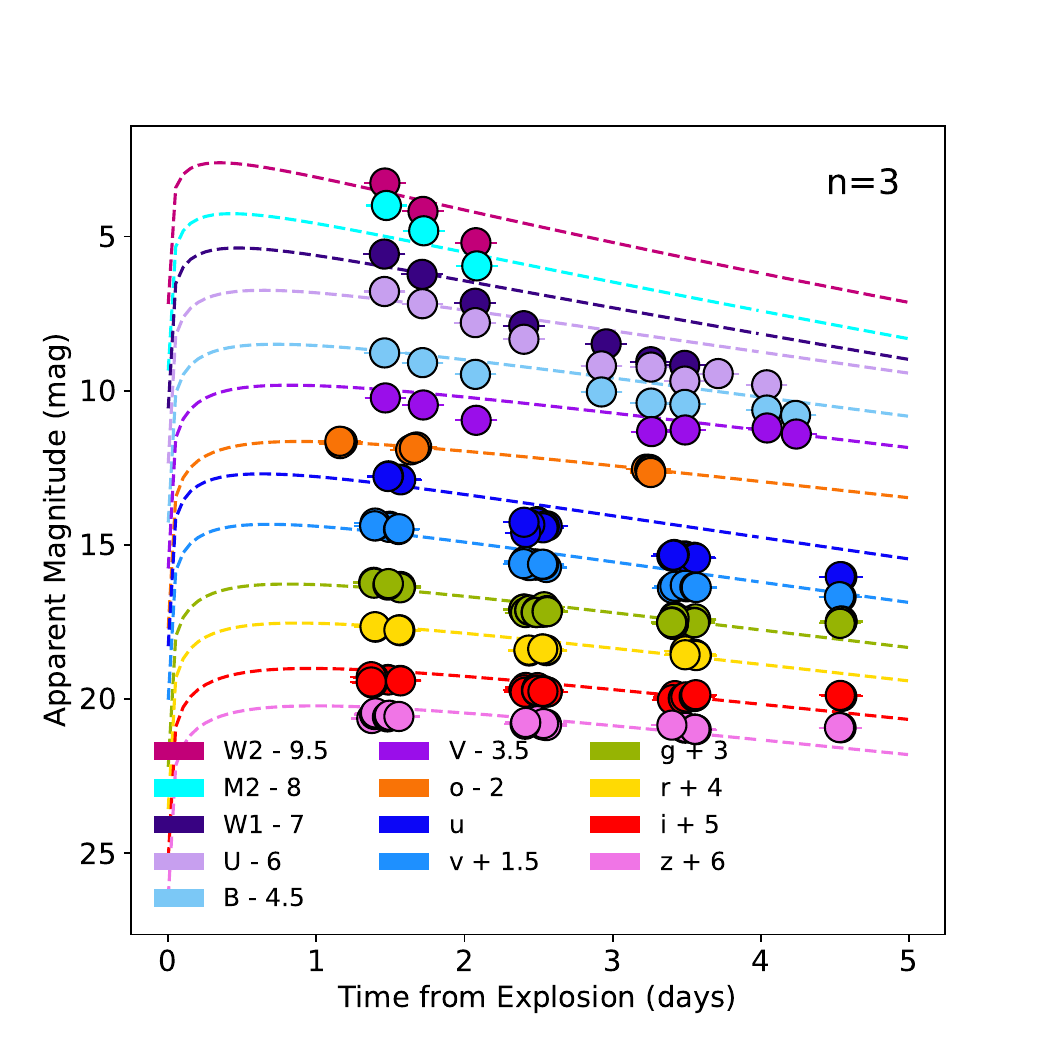}
\includegraphics[scale = 0.5]{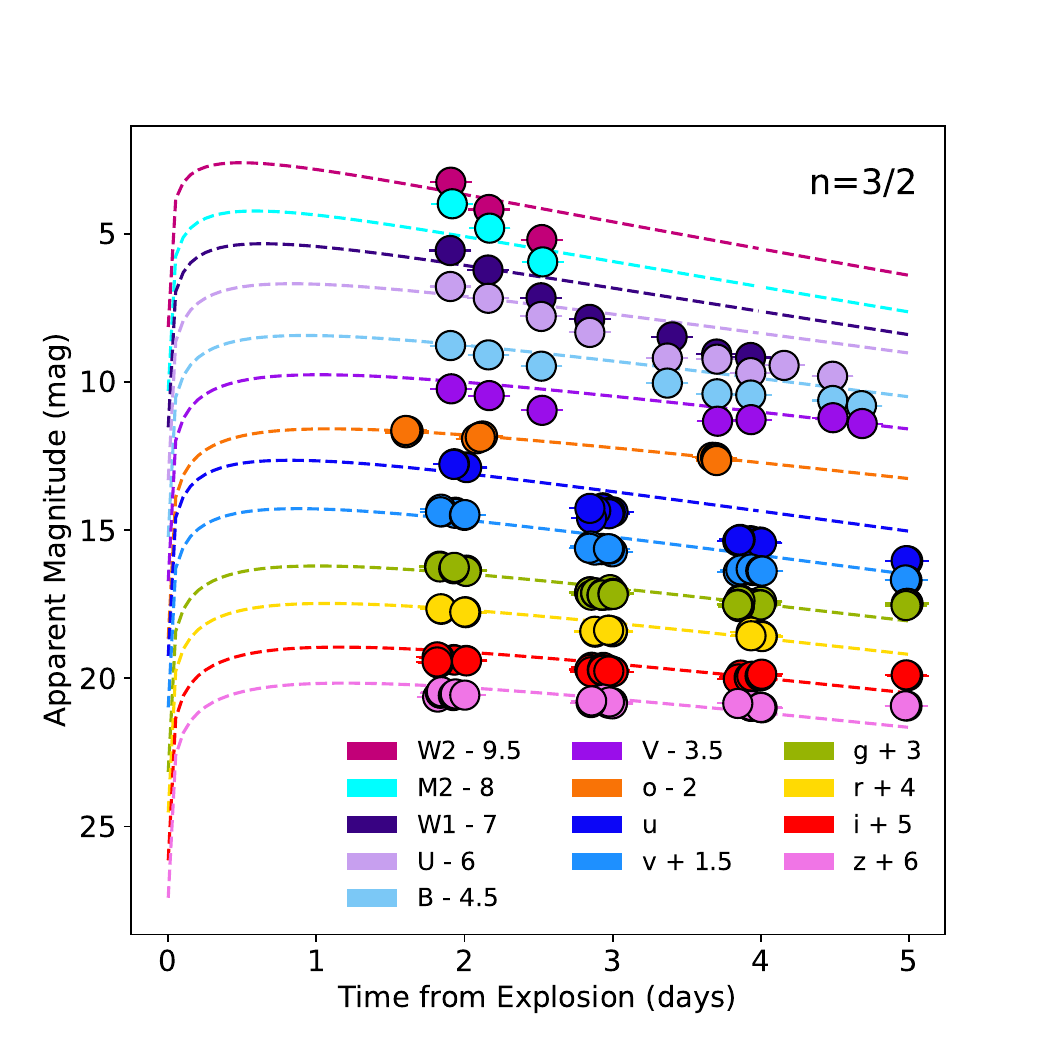}
\caption{Shock cooling model fits to the early ($<$5 days) photometric data points (see Section~\ref{lc-model}). Left and right panels represent the fitting for polytropic indices n = 3 and n = 3/2, respectively. The best-fit parameters are provided in Table~\ref{LC-fitting}. The dotted lines and symbols indicate the best-fitting and multi-band data, respectively.}
\label{Plot-fitting}
\end{figure*}

\section{Light curve modeling and explosion parameters}\label{lc-model}

In double-peaked IIb SNe, the propagating shock breaks out the optically thick progenitor envelope after the SN explosion \citep{1977falk}. The shock-generated electromagnetic radiation powers the first peak, and a fast decline occurs due to the subsequent cooling of the expanding (extended) envelope \citep{Soderberg2012S}. The secondary peak emerges within a few weeks, which is powered by the radioactive decay of $^{56}$Ni synthesized in the explosion. The duration of the shock breakout phase \citep{Waxman2017W} remains very short ($\sim$\,hours) and is therefore usually missed in observations. The mass of the envelope and the density structure of the outer layers characterize the shock cooling phase. Accordingly, various hydrodynamical and analytical models have been proposed to understand the evolution of IIb SNe progenitors and the explosion. It includes helium stars with a thin hydrogen envelope \citep[see][]{Shigeyama1994S, Woosley1994W, Blinnikov1998B, Rabinak-2011, 2014nakar, 2012bersten, Katz2012K, Bersten2018NaturB}.

The extended progenitors and low-mass envelopes of the double-peaked SN~1993J and SN~2011dh were explained with theoretical modeling \citep{2014nakar}. These calculations were successively improved in later studies by incorporating a two-zone model of variable power-law density for better modeling of the early shock cooling features \citep[see][]{Piro2021P}. Further, \citet{Sapir2017S} refined the shock cooling model, considering a complicated polytropic density profile, which was missing in previous models. Recently, \citet{Morag-2023} proposed another model for similar studies by combining previous model solutions of \citet{Sapir2013ApJ-S} and \citet{Sapir2017S}. 

Both models have some limitations and approximations. Namely, the temperature validity is only for $\ge$0.7 eV $\sim$8120 K for both models. Also, the valid envelope mass range is 0.1 $M_{\sun} \ge M_{e}$ $\ge 10 M_{\sun}$ and 1 $M_{\sun} \ge M_{e}$ $\ge$ 10 $M_{\sun}$ for \citet{Sapir2017S} and \citet{Morag-2023}, respectively. Here, it is emphasized that the \citet{Sapir2017S} model is preferred over \citet{Morag-2023} as the previous is more suitable for SNe IIb due to a better $M_{e}$ acceptable range \citep[but see][]{Farah2025F}. Therefore, we use the \citet{Sapir2013ApJ-S} model in our further analysis, which is implemented in the open-source package \texttt{shock-cooling-curve} \citep{Venkatraman2024V}. The code follows a Markov Chain Monte Carlo (MCMC) routine. Considering that the temperature criteria (see Fig.~\ref{fig:temp-radius}, top panel) are satisfied up to less than 5 days of the shock cooling phase, and hence, the model was fitted to this phase only.

Both polytropic indices n = 3 and n = 3/2, which signify the radiative and convective envelopes, were executed. The MCMC routine was operated to achieve the best fit major parameters such as envelope mass (M$_{en}$), progenitor radius (R), and shock velocity scale (v$_{sh}$). It also estimates the explosion time (t$_0$). The best-fitted parameters were achieved after numerous runs (2000), and the resulting model is plotted in Fig.~\ref{Plot-fitting}. The estimated parameter R, M$_{en}$, v$_{sh}$ and t$_{0}$ values are: 169.19 R$_{\sun}$, 0.24 M$_{\sun}$, 2.35 $\times$ 10$^{9}$ cm s$^{-1}$ and --\,0.55 days (for polytropic index n = 3); and 200.35 R$_{\sun}$, 0.03 M$_{\sun}$, 1.55 $\times$ 10$^{9}$ cm s$^{-1}$ and --\,1.00 days (for polytropic index n = 3/2), respectively. These are listed in Table~\ref{LC-fitting}. The corner plots resulting from the fitting are shown in Appendix~\ref{Apendix} (Fig.~\ref{Fitting-A1} and \ref{Fitting-A2}).

We used bolometric light curves computed in Section~\ref{lc-Bol} to estimate the explosion parameters of SN~2024aecx. The conceptualizations given by \citep{1982arnett} and \citep{2008valenti} were used to fit the second peak, which is mainly believed to be powered by the radioactive decay of $^{56}$Ni. The models are based on several assumptions that the ejecta are spherically symmetric and homologous, the opacity ($\kappa_{\rm opt}$) is constant, and $^{56}$Ni is centrally located and unmixed \citep{1982arnett, 2008valenti, Cao2013C}. The parameters $M_{\rm Ni}$ and $\tau_{\rm m}$ are the nickel mass and diffusion time-scale, respectively, and were kept as free variables in the fitting. The ejecta kinetic energy $E_{\rm k}$ and $\tau_{\rm m}$ are represented by
\begin{equation}\label{eq_1}
\tau_{\rm m} = \sqrt{2} \left( \frac{\kappa_{\rm opt}}{\beta c} \right)^{1/2} \left( \frac{M_{\rm ej}}{v_{\rm ph}} \right)^{1/2},
\end{equation}

\begin{equation}\label{eq_2}
E_{\rm KE} \approx \frac{3}{5}\frac{M_{\rm ej}v^2_{\rm ph}}{2},
\end{equation}

Here, $\beta \approx 13.8$ is a constant of integration \citep{2008valenti}, v$_{\rm ph}$ is photospheric velocity, and $c$ is the speed of light. The optical opacity $\kappa_{\rm opt}$ is a constant \citep{Chugai797C, 2016ApJ...818...79T, 2018AA...609A.136T}. The fits were performed with the transient fitting model implemented in \texttt{Redback} \citep{Sarin2024MNRASS}. In the fitting, the ejecta mass $M_{ej}$, fraction of nickel f$_{Ni}$, representative initial ejecta velocity v$_{ej}$, $\kappa_{\rm opt}$, the gamma-ray opacity $\kappa_{\rm \gamma}$ were kept as free parameters. The estimated values of the explosion parameters are: $M_{Ni}$ = 0.15 $\pm$ 0.06 M$_{\sun}$, $M_{ej}$ = 0.70$^{+0.18}_{-0.16}$ M$_{\sun}$ and $E_{\rm KE}$ = 0.16 $\pm$ ${0.05}$ $\times$ 10$^{51}$ erg. $E_{\rm KE}$ was estimated using the above equation, and the photospheric velocity was taken as 6000 km s$^{-1}$ (see Section~\ref{spec-prop}).

\section{Discussion and Summary}\label{summary}

The early phase multi-band observations of double-peaked IIb SNe with a good cadence are limited in the literature. Existing large survey facilities are restricted to a single or double band on a particular night. The temporal evolution of the shock breakout, followed by the shock cooling, is rapid, and the duration depends on the progenitor structure. Therefore, observations in the early phase after the explosion are challenging. Because of their peculiar observational properties, SNe IIb are regarded as gap-bridging events between IIP and Ib/c. Hence, additional study of such SNe can play a crucial role in understanding stellar evolution and supernova physics. We investigated simultaneous multi-band imaging of SN~2024aecx, including the shock cooling phase. Our analysis infers that the duration of the shock cooling phase is longer in bluer bands, but the rise time to the secondary peak (after the explosion) is shorter in those bands. Also, the second peak is broader in redder bands. It is further noted that the shock cooling phase of SN~2024aecx is shorter than several Type IIb events (SNe~1993J, 2011fu, and 2013df). 

\begin{figure}
\centering
\includegraphics[width=\columnwidth]{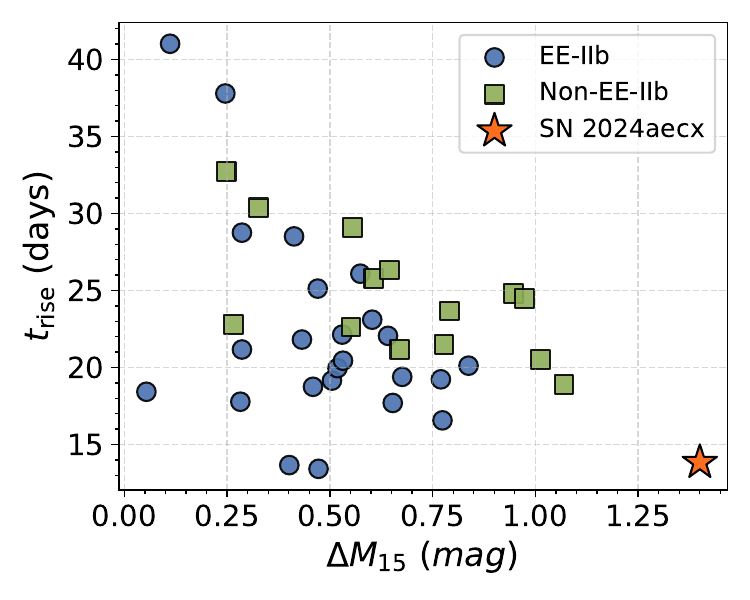}
\caption{The $\Delta$m$_{15}$ (ATLAS) value for SN~2024aecx estimated in Section~\ref{lc-prop} is compared with the sample from \citet{Ayala2025A}. Here, EE-IIb and Non-EE-IIb indicate Type IIb SNe with an early flux excess and without early flux excess, respectively \citep[for details, see][]{Ayala2025A}. SN~2024aecx is indicated with a star symbol.}
\label{fig:deltam}
\end{figure}

Recently, \citet{Ayala2025A} performed a detailed statistical study of early light curves of IIb SNe based on a sample from the ATLAS survey and well-studied literature data. They categorize these events into two groups: EE (an early flux excess SNe followed by a secondary peak) and non-EE (without early flux excess SNe). SN~2024aecx falls to their EE group (cf. double peak). In Fig.~\ref{fig:deltam}, the t$_{rise}$ (rise time to secondary peak after the explosion) and $\Delta$m$_{15}$ of their sample and the values estimated for SN~2024aecx are displayed (also see Table~\ref{tab_lc_p}). There is a clear trend that events with faster post-peak decline rates attain a brighter secondary peak earlier \citep{Ayala2025A}. It is worth noting that SN~2024aecx has very fast evolution among the sample, which is also evident from Figure~\ref{fig:fig_abs}. Nevertheless, such an inference can be confirmed with an additional data sample in the future. Further, the shock cooling duration of SN~2024aecx (see Table~\ref{tab_lc_p}) is significantly less than the mean duration (8.85 days) as estimated in \citet{Ayala2025A}. With a peak absolute magnitude (second peak) of M$_{g}$ = --17.94 mag, SN~2024aecx is one of the brightest events in the SNe IIb sample, but it does not follow the trend that brighter SNe exhibit longer rise times; rather, it peaked earlier. 
Similarly, the color evolution of SN~2024aecx is unique, and various colors display a trend of blue-red and red-blue-red between $\sim$\,0\,--\,8 days and $\sim$\,8\,--\,40 days. Such a feature is likely because of the interplay between the shock cooling and the cooling of the expanding ejecta. 
During the monitoring of NGC~3521 for more than a year with Mephisto, the forced photometry at the SN location does not exhibit any reliable flux enhancement.

The spectroscopic properties of SN~2024aecx have similarities with typical IIb SNe, but the hydrogen line features are weak and could only be seen up to about 1 month after the explosion. To discern several of the spectral features, we performed a crude spectral modeling of SN~2024aecx using \texttt{SYNAPPS}. A significant number of lines were identified and corroborated, justifying the usual spectral properties of IIb events (see Section~\ref{Synapps}). However, a few spectral lines remain evasive. The presence of the hydrogen line in the early spectra of SN~2024aecx is an important clue to differentiate it from Type Ib events. However, the possible contamination arising due to the blending of \ion{Si}{2} line near 6500 \AA\ can not be ignored.

Another important line \ion{He}{1} 5876 \AA, evolved from the beginning in SN~2024aecx against SN~2013df, where it started appearing $\sim$10 days post-explosion. Nevertheless, the trough near this line could be blended with \ion{Na}{1} (5890 \AA, 5986 \AA). In Type Ic SNe, the identification of \ion{He}{1} 5876 \AA\, line is debatable \citep{Clocchiatti1996C, Matheson2001M, Elmhamdi2006E, Sauer2006S, Dessart2015D, Milisavljevic2015M}. \citet{Liu2016ApJL} analyzed a large optical spectroscopic data of Ib and Ic SNe. In their study, no convincing signs of either \ion{He}{1} 6678 \AA\ or 7065 \AA\ were noticed in the mean spectra of Ic SNe. They suggested no obvious \ion{He}{1} lines in Ic SNe \citep[see also][]{Modjaz2016M}. The \ion{He}{1} 10830 \AA\ and 20581 \AA\ lines in the NIR region are indeed useful for an affirmation between the helium-rich and helium-poor events \citep{Shahbandeh2022S}. However, we can not comment on this without NIR spectra of SN~2024aecx. It is worth mentioning that there are limitations with the spectral modeling presented in this paper. A thorough analysis of the ejecta composition and evolution could be carried out using more sophisticated spectral modeling codes (e.g., CMFGEN, TARDIS, etc.), which is beyond the scope of this work.

The shock cooling phase multi-band data ($<$ 5 days) of SN~2024aecx were modeled with \citet{Sapir2017S} shock cooling analytical model. The radiative (n = 3) and convective (n = 3/2) polytropic indices were used to infer the progenitor properties (see Section~\ref{lc-model}). The estimated best-fit envelope mass has a range of 0.03\,--\,0.24 $M_{\sun}$, indicating a moderately stripped progenitor. This is similar to SN~1993J progenitor, where M$_{e}$ was estimated less than 0.4 $M_{\sun}$ \citep{Woosley1994W}. Such a smaller envelope mass also corroborates the weak hydrogen appearance in SN~2024aecx spectra. Furthermore, the progenitor radius is $\sim$169 R$_{\sun}$ and $\sim$\,200 R$_{\sun}$ for n = 3 and 3/2, respectively. This range has a similarity with progenitor radii of SN~2016gkg \citep{Tartaglia2017T, Piro2017ApJ...846...94P}. The overall observational properties and analytical modeling parameters of SN~2024aecx indicates that its progenitor was possibly less extended than SN~1993J \citep[$\sim$\,600 R$_{\sun}$][]{Maund2004M}, SN~2011fu \citep[$\sim$450 R$_{\sun}$][]{Kumar2013K, 2015morales} and SN~2013df \citep[$\sim$550 R$_{\sun}$][]{VanDyk2014V}. However, \citet{2014morales} estimated a smaller radius for SN~2013df (64\,--\,169 R$_{\sun}$). The intermediate extended progenitor of SN~204aecx was also favored, considering a shorter shock cooling time of SN~2024aecx (see Section~\ref{tab_lc_p}). Nevertheless, a more robust hydrodynamical modeling can provide a more clear picture of the SN~2024aecx progenitor system.

The modeling of pseudo-bolometric light curve implies an ejecta mass of $\sim$\,0.70 M$_{\sun}$, an explosion energy $\sim$\,0.16 $\times$ 10$^{51}$ erg, and the synthesized $^{56}$Ni $\sim$\,0.15 M$_{\sun}$. The production of $^{56}$Ni in SE-SNe is correlated with the explosion properties and the structure of the progenitor core \citep[see][and references therein]{Suwa2019S}. Higher $^{56}$Ni mass signifies a more luminous event, which supports the brighter peak of SN~2024aecx. However, the $^{56}$Ni evaluation using Arnett's rule may overestimate the $^{56}$Ni value in SE-SNe as Arnett's model is more appropriate for Type Ia SNe \citep{2015MNRAS.453.2189D, 2016MNRAS.458.1618D, 2019ApJ...878...56K}. Recent sample-based studies suggest $^{56}$Ni production range of 0.060 -- 0.124 for Type IIb SNe \citep{Anderson2019A, 2019MNRAS.485.1559P, Afsariardchi2021A, 2023ApJ...955...71R}. 

The estimated $^{56}$Ni value for SN~2024aecx is consistent with \citet{Anderson2019A} and also with the derived $^{56}$Ni values for SN~2011fu \citep{Kumar2013K} and SN~2013df \citep{2014morales}. The light curve shape of SE-SNe is also influenced by the mixing of $^{56}$Ni in the ejecta \citep{1988ApJ...333..754E, 1997ASIC..486..821W, 1990ApJ...361L..23S, 2014MNRAS.438.2924C, 2015MNRAS.453.2189D, 2016MNRAS.458.1618D}. The $^{56}$Ni mixing may arise due to various reasons, e.g., large-scale asymmetric explosion jets and asymmetry of the propagating shock, etc. \citep{Maund2009, 2012MNRAS.424.2139D}. The early color evolution can suggest the degree of mixing. The monotonic and non-monotonic color evolution hints at strong and weak mixing, respectively \citep{2019ApJ...872..174Y, 2020MNRAS.497.1619M}. SN~2024aecx early colors, which display a non-monotonic evolution, may be attributed to a weak to moderate $^{56}$Ni mixing. Further, as the ejecta mass of SN~2024aecx is smaller, it will result shorter diffusion time of the radiation \citep{2023ApJ...955...71R} and so faster light curve evolution.

\section{Software and third party data repository citations} \label{sec:soft}

\software{astropy \citep{astropy:2022}, pandas \citep{reback2020pandas}, numpy \citep{harris2020array}, scipy \citep{2020SciPy-NMeth}, Jupyter-notebook \citep{Kluyver2016jupyter}, SWarp \citep{Bertin-2002ASPC..281..228B}, SExtractor \citep{Bertin-1996A&AS..117..393B}, Light Curve Fitting package \citep{2023zndo...8049154H}}.

\section*{Acknowledgments}
Mephisto is developed at and operated by the South-Western Institute for Astronomy Research of Yunnan University (SWIFAR-YNU), funded by the ``Yunnan University Development Plan for World-Class University" and ``Yunnan University Development Plan for World-Class Astronomy Discipline". The authors acknowledge support from the ``Science \& Technology Champion Project'' (202005AB160002) and from two ``Team Projects" -- the ``Top Team'' (202305AT350002) and the ``Innovation Team'' (202105AE160021), all funded by the ``Yunnan Revitalization Talent Support Program". We also acknowledge the National Key Research and Development Program of China (2024YFA1611603). BK is supported by the ``Special Project for High-End Foreign Experts", Xingdian Funding from Yunnan Province. JZ is supported by the National Natural Science Foundation of China (12173082, 12333008), the Yunnan Fundamental Research Projects (grants 202501AV070012, 202401BC070007 and 202201AT070069), the Top notch Young Talents Program of Yunnan Province, the Light of West China Program provided by the Chinese Academy of Sciences, the International Centre of Supernovae, Yunnan Key Laboratory (No. 202302AN360001). 

We thank the staff of IAO, Hanle, and CREST, Hosakote, that made HCT observations possible. The facilities at IAO and CREST are operated by the Indian Institute of Astrophysics, Bangalore. We acknowledge the support of the staff of the LJT. Funding for the LJT has been provided by the CAS and the People's Government of Yunnan Province. The LJT is jointly operated and administrated by YNAO and the Center for Astronomical Mega-Science, CAS. We also acknowledge WISeREP - \url{https://www.wiserep.org}. 

This work has made use of data from the Asteroid Terrestrial-impact Last Alert System (ATLAS) project. The Asteroid Terrestrial-impact Last Alert System (ATLAS) project is primarily funded to search for near earth asteroids through NASA grants NN12AR55G, 80NSSC18K0284, and 80NSSC18K1575; byproducts of the NEO search include images and catalogs from the survey area. This work was partially funded by Kepler/K2 grant J1944/80NSSC19K0112 and HST GO-15889, and STFC grants ST/T000198/1 and ST/S006109/1. The ATLAS science products have been made possible through the contributions of the University of Hawaii Institute for Astronomy, the Queen's University Belfast, the Space Telescope Science Institute, the South African Astronomical Observatory, and The Millennium Institute of Astrophysics (MAS), Chile. We also thank the referee for insightful comments and suggestions that significantly improved the manuscript.
\bibliography{2-SN2024aecx}
\bibliographystyle{aasjournalv7}

\appendix \label{Apendix}

\begin{figure}
\centering
\includegraphics[scale=0.65]{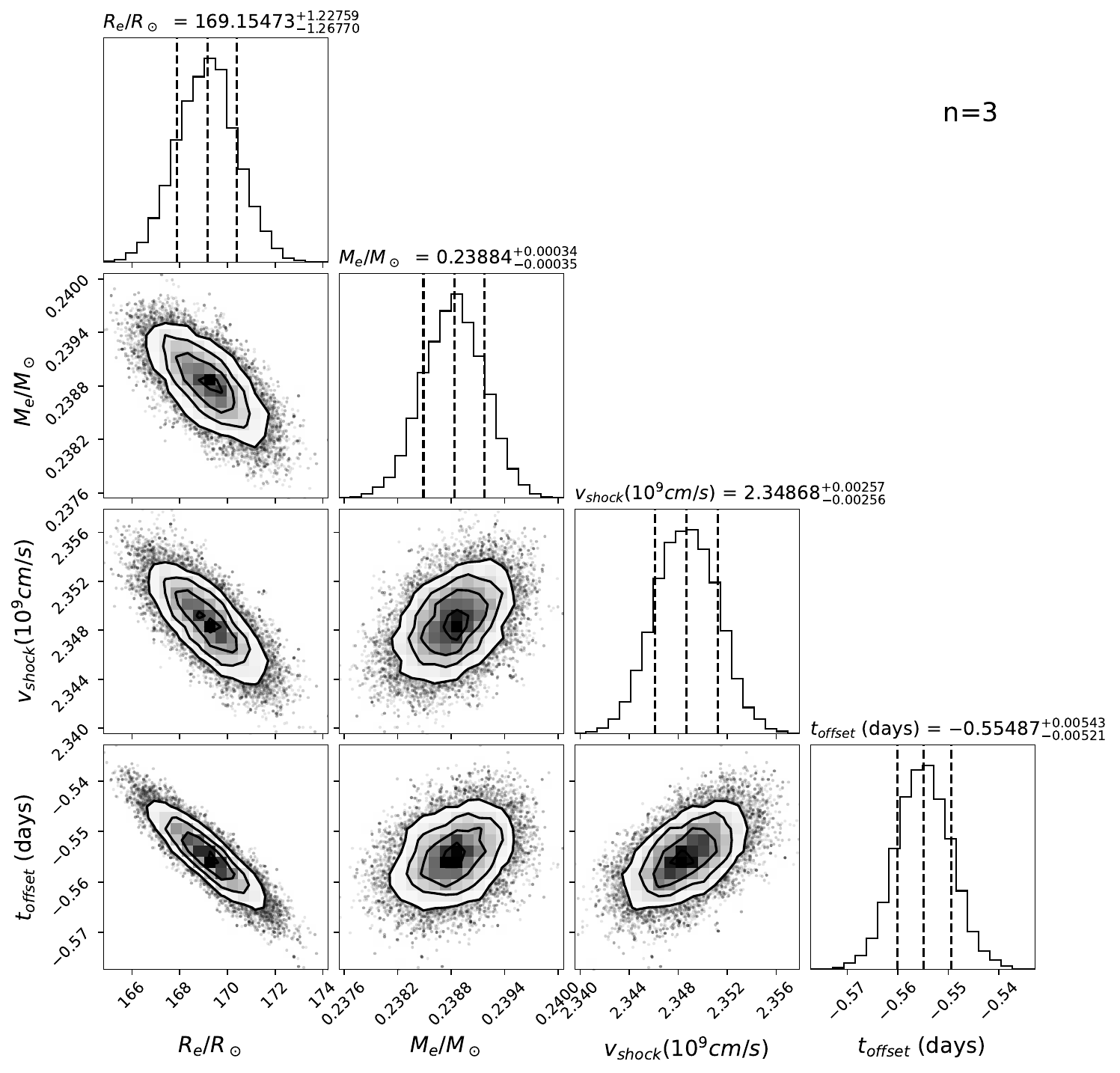}
\caption{The corner plot for the analytical fitting as described in Section~\ref{LC-fitting}. The indicated parameter distribution is for polytropic index n = 3.}
\label{Fitting-A1}
\end{figure}

\begin{figure}
\centering
\includegraphics[scale=0.65]{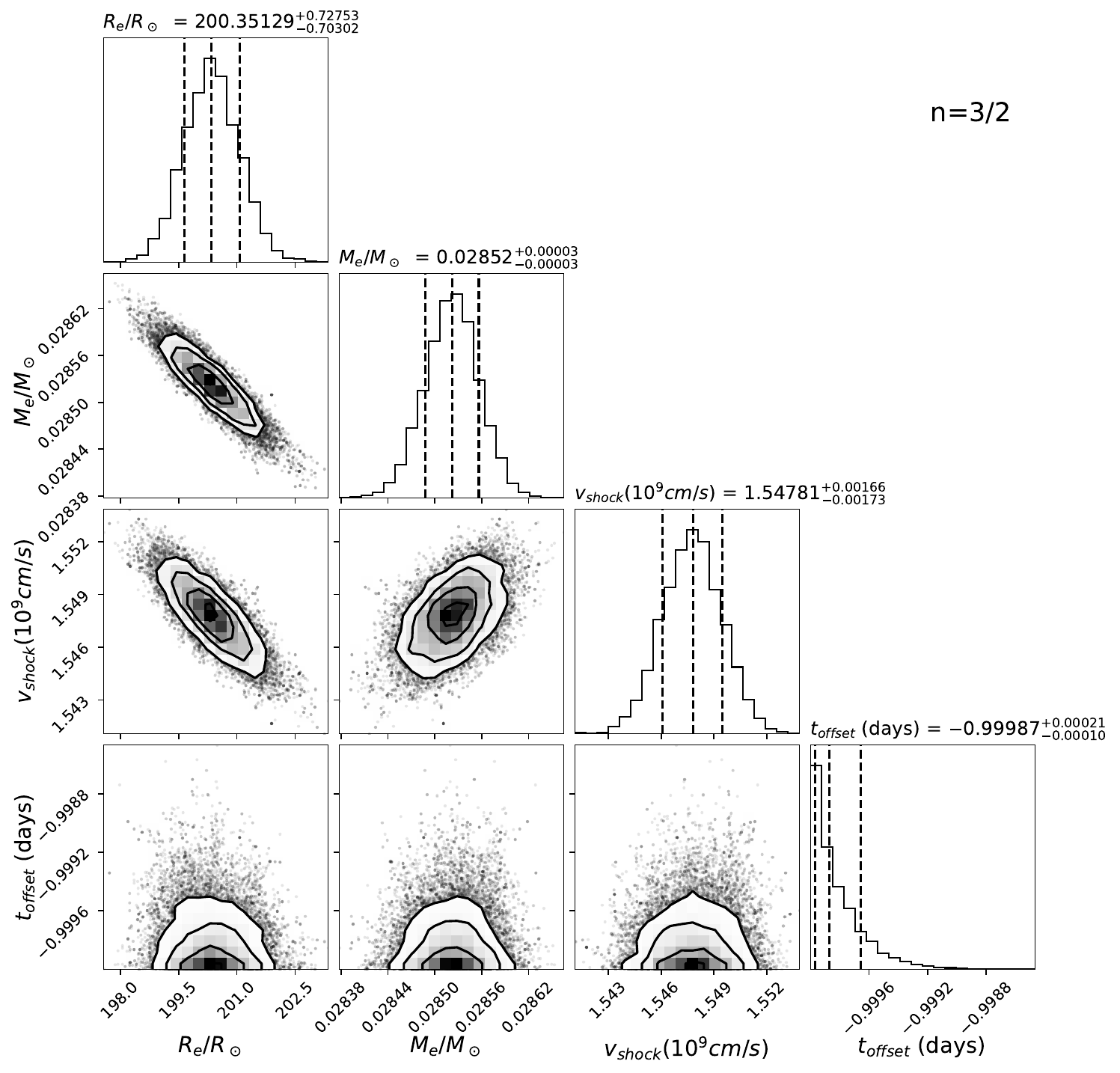}
\caption{Similar to Fig.~\ref{Fitting-A1} but for polytropic index n = 3/2.}
\label{Fitting-A2}
\end{figure}

Fig.~\ref{Fitting-A1} and Fig.~\ref{Fitting-A2}, represent the corner plots for two polytropic indices n = 3 and n = 3/2 (see Section~\ref{lc-model} for more details).

\end{document}